\documentstyle[epsfig]{mn}

\newcommand{\mnras}{MNRAS}
\newcommand{\apj}{ApJ}

\newcommand{\bc}{\begin{center}}
\newcommand{\ec}{\end{center}}

\title[Early structure in $\Lambda$CDM]
{Early structure in $\Lambda$CDM}

\author[L.~Gao, et al.]  {L.~Gao$^1$\thanks{Email:
        gaoliang@mpa-garching.mpg.de}, S.~D.~M.~White$^1$,
        A.~Jenkins$^2$, C.~S.~Frenk$^2$, Volker Springel$^1$ \\
        $^1$Max--Planck--Institut f\"ur Astrophysik, D-85748 Garching,
        Germany \\ $^2$Institute for Computational Cosmology,
        Department of Physics, University of Durham,South Road, Durham
        DH1 3LE,U.K.}

\begin{document}
\label{firstpage} \maketitle

\begin{abstract}
We use a novel technique to simulate the growth of one of the most massive
progenitors of a supercluster region from redshift $z\sim 80$, when its mass
was about $10{\rm M_{\odot}}$, until the present day. Our nested sequence of
$N$--body resimulations allows us to study in detail the structure both of the
dark matter object itself and of its environment. Our effective resolution is
optimal at redshifts of 49, 29, 12, 5 and 0 when the dominant object has mass
$1.2\times 10^5$, $5\times 10^7$, $2\times 10^{10}$, $3\times 10^{12}$ and
$8\times 10^{14}h^{-1}{\rm M_\odot}$ respectively, and contains $\sim 10^6$
simulation particles within its virial radius. Extended Press-Schechter theory
correctly predicts both this rapid growth and the substantial overabundance of
massive haloes we find at early times in regions surrounding the dominant
object. Although the large-scale structure in these regions differs
dramatically from a scaled version of its present-day counterpart, the
internal structure of the dominant object is remarkably similar. Molecular
hydrogen cooling could start as early as $z\sim 49$ in this object, while
cooling by atomic hydrogen becomes effective at $z\sim 39$. If the first stars
formed in haloes with virial temperature $\sim 2000$K, their comoving
abundance at $z=49$ should be similar to that of dwarf galaxies today, while
their comoving correlation length should be $\sim 2.5h^{-1}$Mpc.
\end{abstract}

\begin{keywords}
methods: N-body simulations -- methods: numerical --dark matter --
galaxies: haloes -- cosmology:theory-early Universe
\end{keywords}
\title{Early structure in $\Lambda$CDM}

\section{Introduction}
Within the cold dark matter (CDM) paradigm for structure formation, first
light in the Universe is usually assumed to have come from stars which
collapsed early at the centres of rare and unusually massive dark matter
haloes associated with high peaks of the initial gaussian overdensity
field. Soon after their birth, these stars began to influence the structure,
the thermodynamics and the chemical content of surrounding gas. The epoch when
this occurred and the precise details of how it happened are not yet well
understood.

Recent simulations including gravitational, chemical and radiative processes
have suggested that metal-free gas in dark matter haloes of virial temperature
${\rm T_{vir} \sim 2000 K}$ and mass ${\rm M} \sim 10^6{\rm M_{\odot}}$ cooled
efficiently by emission from molecular hydrogen and collapsed to form a star
at redshifts $18<z<30$.  (Abel et al. 1998, 2002; Fuller \& Couchman 2000,
Bromm et al. 2002; Yoshida et al. 2003; for reviews see Bromm \& Larson 2004
and Glover 2004). To achieve the required resolution, these calculations
followed evolution within very small cubic subvolumes of the Universe assuming
periodic boundary conditions.  Such boundary conditions suppress all
fluctuations with wavelength longer than the side of the simulated region. As
we shall see, this can have drastic effects, particularly on the morphology of
large-scale structure and on the abundance and formation history of objects
with mass more than roughly $10^{-4}$ of the total mass in the simulated
region.

In fact, the first haloes of any given mass (and thus quite probably the first
stars) do not form in ``typical'' regions at all, but rather in
``protocluster'' regions (White \& Springel 2000; Barkana \& Loeb 2004).  This
is an effect of the relatively large contribution from Mpc wavelengths to the
fluctuation amplitude on even the smallest mass scales, and is reflected, for
example, in the plots presented by Mo \& White (2002) for the spatial
clustering strength of haloes for redshifts out to $z\sim 20$.  Conclusions
about early objects reached from simulations of small and ``typical'' regions
may thus be misleading; in particular, the formation redshift of the first
haloes of any given mass or temperature will be systematically
underestimated. An alternative strategy is to simulate regions which are
constrained to have high overdensity or to contain a high amplitude density
peak (Abel et al. 1998, Fuller \&~Couchman 2000; Bromm et al.  2002) but this
leads to interpretational difficulties because there is no clear
relation between the statistical properties of the resulting objects and those
of the rare objects of similar mass which would form from unconstrained
initial conditions (White 1996).

The formation of the first stars has also been studied by
combining semi-analytical modelling of baryonic processes (e.g.
Tegmark et al 1997; Hutchings et al. 2002; Yoshida et al. 2003)
with Press-Schechter or extended Press-Schechter predictions of
the number density of dark matter haloes (Press \& Schechter 1974;
Bond et al. 1991; Bower 1991; Lacey \& Cole 1993, 1994). Over the
limited mass range that they test, recent N-body simulations have
shown that the halo number density in the concordance cosmology is
reasonably well matched by the Sheth \& Tormen (1999) variant of
the Press-Schechter model (Jenkins et al. 2001; Reed et al. 2003;
Yahagi et al. 2004; Gao et al. 2004a, Springel et al. 2005).
Unfortunately, this work is still far from testing the regime that
is relevant for studies of the first stars.

In this paper, we attempt to simulate an example of one of the first haloes
capable of forming stars. Anticipating that these will be progenitors of
massive present-day objects, we perform a series of nested resimulations of
the most massive progenitor of a rich cluster and its supercluster
environment. In each resimulation we increase the resolution, we include
additional small-scale structure in the initial conditions, and we start at
higher redshift in order to capture the early phases of nonlinear
evolution. As anticipated from the above discussion, we find that our
simulated halo would likely form its first star at significantly higher
redshifts than suggested by previous work. In this paper, however, we simulate
only the formation and evolution of the dark halo itself.  A more detailed
consideration of baryonic processes is deferred to a companion paper (Reed et
al. 2005c) and to later papers which include simulation of hydrodynamics and
radiative cooling.

Our paper is structured as follows. In Section~2, we describe our simulation
sequence, the methods used to set up initial conditions, to integrate the
equations of motion, and to identify haloes, and the evolution in mass and
effective abundance of the dominant object.  In Section~3, we present images
of our main halo and its immediate surroundings at $z=49, 29, 12, 5$ and
$0$. We also present density profiles and an analysis of substructure for this
halo at these same times, and we chart its remarkably rapid growth in virial
temperature. Internal structure is surprisingly similar at different times,
whereas halo environment changes systematically and qualititively with
increasing redshift. In Section~4 we extend our study of environment to larger
scales, highlighting the qualitative difference in structure between early and
late times.  At redshift 50 the dark matter distribution in our simulation is
very different from a scaled copy of that at $z=0$. In this section we also
compare the number density of haloes in our simulations with analytical
models, showing that there is an extremely strong clustering bias which is
quite well modelled by a suitable version of extended Press-Schechter theory.
In Section~5, we discuss the some possible implications of our calculations
for the formation on the first stars. Finally, we summarize our conclusions in
Section~6.

\section{Simulations details}
We adopt standard values for the cosmological parameters. For the
mean densities of dark matter, dark energy and baryons (in units
of the critical density) we take $\Omega_{\rm DM}=0.26$,
$\Omega_\Lambda=0.7$ and $\Omega_{\rm B}=0.04$ respectively. The
present value of the Hubble parameter is set to $h=0.7$ and the
extrapolated linear amplitude of fluctuations in a sphere of
radius $8 h^{-1}{\rm M}_\odot$ is set to $\sigma_{8}=0.9$. We
computed the initial linear power spectrum down to $k\sim
2000h{\rm Mpc}^{-1}$ using {\small CMBFAST} with a primordial
spectral index $n=1$ (Seljak \& Zaldarriaga, 1996). It was
necessary to extrapolate a further order of magnitude in
wavenumber to reach the Nyquist frequency corresponding to our
highest resolution resimulation. This was done using a power-law
matched to the slope ($dn/dlnk=-2.99$) at the join. Strictly
speaking, there should be some curvature, but at these wavenumbers
the slope is very close to its asymptotic value of~$-3$.

\subsection{The simulations}

\begin{table*}
\caption{Numerical parameters for the R series simulations}
\begin{center}
\begin{tabular}{l c c c c c}
   \hline
    & $R1$  & $R2$  & $R3$ & $R4$ & $R5$\\
   \hline
    $N_p$    &8457516 &5804755 &8658025  &41226712 &73744737\\
    $m_p[h^{-1}{\rm M_\odot}]$ & $5.12\times 10^{8}$  & $2.2\times 10^{6}$ &
    $1.24\times 10^{4}$ & $29.5$ &$0.545$\\
    $\epsilon[h^{-1}{\rm kpc}]$ &5.0  &0.8 & 0.15 & 0.017 &0.0048\\
    $M_{200}[h^{-1}{\rm M_\odot}]$ &$8.1\times 10^{14}$ &$4.6\times 10^{12}$
    &$2.0\times 10^{10}$,&$5.2 \times 10^7$ &$1.2\times 10^5$\\
    $r_{200}[h^{-1}{\rm kpc}]$ &1514.1 &401.5 &65.9 &9.1 & 1.2\\
    $z_{i}$ &39 &149 &249 &399 &599 \\
    $z_{f}$ &0.0  &5.0 &12.04  &29.04 &48.84 \\
    \hline
\end{tabular}
\end{center}
\end{table*}

It is a challenge to simulate the formation of early structure in
a CDM universe because the first collapsed objects are very small.
In addition, the slope of the matter power spectrum at high
wavenumber approaches the critical value~$-3$ for which equal
contributions to the variance of the density field come from each
logarithmic interval in wavenumber.  Thus a very wide range of
scales must be included in order to simulate correctly the early
formation of small objects. The required dynamic range is beyond
present algorithms and computer resources by a wide margin so it
is currently impossible to carry out simulations with sufficient
resolution everwhere to follow the nonlinear dynamics of the first
collapsing structures.

We have devised a multigrid procedure which circumvents this
problem and allows us to follow the growth of an object in its
full cosmological context from $z=80$ to $z=0$. Over this period
its mass increases by about 13 orders of magnitude. The procedure
works as follows:

\begin{description}
\item[(1)] We identify a rich cluster halo at $z=0$ in a cosmological
simulation of a very large volume. \item[(2)] We resimulate the evolution of
this rich cluster and its environment with much higher mass and force
resolution, while following the remainder of the simulation volume at lower
resolution than before. We ensure the proper treatment of small-scale
structure by starting the resimulation at higher redshift than the original,
and including in its initial conditions a random realisation of the
fluctuations at wavenumbers between the Nyquist frequencies of the original
and the higher resolution particle distributions. \item[(3)] We find the most
massive object in the high resolution region of this resimulation at each
redshift and identify the time when it {\it first} contains more than about
10000 particles within its virial radius. \item[(4)] We then resimulate the
evolution of this progenitor object and its immediate surroundings with
further improved mass and force resolution, while again following the more
distant matter distribution at lower resolution. \item[(5)] We iterate steps~3
and~4 until we reach the desired redshift and progenitor mass.
\end{description}

In practice, we chose at random a rich cluster of virial mass $8\times
10^{14}h^{-1}{\rm M_\odot}$ from the VLS simulation of the Virgo consortium, a
cosmological simulation of a cubic region of side $479h^{-1}$Mpc (Jenkins et
al. 2001, Yoshida et al. 2001).  The first resimulation of this cluster
involved a single application of the ``zoom'' technique first introduced by
Navarro \& White (1994) in the implementation described and tested by Power~et
al. (2003). This was one of the cluster resimulations analyzed in Navarro et
al. (2004) and Gao et al.  (2004a,b,c). For the present paper we repeated this
``zoom'' step four more times reaching ever higher mass and force
resolution. We thus have a total of 5 resimulations in addition to the
original VLS simulation. We label these R1 ($z_f=0$), R2 ($z_f=5$),
R3($z_f=12$), R4 ($z_f=29$) and R5 ($z_f=49$), where $z_f$ is both the final
redshift of each resimulation and the redshift at which the most massive
progenitor in the high-resolution region of the previous resimulation was
identified.  We checked the virial mass of the main object and the morphology
of the surrounding structure between each pair of simulations at these times,
finding excellent agreement in all cases. The first object to reach a mass
greater than 10000 particles in the high-resolution region of R1 turned out
not to be a progenitor of the main cluster itself, but rather of a smaller
$\sim 10^{14}h^{-1}{\rm M_\odot}$ member of the same $z=0$ supercluster. At
$z=5$, this object was $15$ per cent more massive than the most massive
progenitor of the main cluster. The objects resimulated in R3, R4 and R5 did,
however, all turn out to be progenitors of this same object.

The high resolution regions of the first three resimulations, R1--R3, had
final radii about $4$ times the virial radius of the final massive object. For
R4 and R5 we chose to follow a more extended region in order to investigate
the large-scale environment of the principal halo. The first three
resimulations used the original periodic boundaries of the parent VLS
simulation. For R4 and R5 we used isolated boundary conditions with sharp
spherical cuts at comoving radii of $5h^{-1}$Mpc and $1.25h^{-1}$Mpc
respectively. At the relevant redshifts the universe is quite homogeneous on
these scales, and our target objects were almost unaffected by the omission of
more distant structure. Note that this cut-off, unlike the imposition of
periodic boundary conditions, does not exclude the contribution to the density
field due to long wavelength modes, although it does significantly affect the
bulk motion of the whole simulated region.  This is of no consequence here.

Further details of our series of resimulations are listed in Table 1. Here
$N_p$ is the total number of particles in the high resolution region of each
resimulation, $m_p$ is the mass of each of these particles, $\epsilon$ is the
gravitational softening parameter (in comoving units), $M_{200}$ is the mass
of the final object within the sphere of radius $r_{200}$ (also given in
comoving units) which encloses a mean overdensity of $200$ relative to
critical, and $z_{i}$ and $z_{f}$ are the initial and final redshifts of each
resimulation.

In most of our resimulations the dominant halo at the final time
has about 2 million particles inside $r_{200}$. Only for R5 at
$z=49$ was this number significantly lower, about $0.2$ million.
For this highest resolution resimulation, the particle mass was
$0.5h^{-1}{\rm M_\odot}$ and the force softening $5h^{-1}$pc in
comoving units. The growth in mass of the principal object we have
followed is shown in the first panel of Figure~\ref{fig:fig1}. (As
noted above, after $z\sim 4$ this is not the most massive object
in the high resolution region of R1.) The rapidity with which the
mass increases at early times is quite remarkable.  Picking a
series of redshifts separated by factors of two in expansion
factor, we find masses of $10^3h^{-1}{\rm M}_\odot$ at $z=63$, of
$2\times 10^7h^{-1}{\rm M}_\odot$ at $z=31$, of $5\times 10^9h^{-1}{\rm
M}_\odot$ at $z=15$, of $4\times 10^{11}h^{-1}{\rm M}_\odot$ at $z=7$,
of $7\times 10^{12}h^{-1}{\rm M}_\odot$ at $z=3$, of $3\times
10^{13}h^{-1}{\rm M}_\odot$ at $z=1$, and of $10^{14}h^{-1}{\rm M}_\odot$ at
$z=0$.

For given halo mass and redshift, one can define a characteristic abundance as
the mean number of haloes of equal or greater mass per unit volume in the
Universe as a whole.  While our iterative procedure is guaranteed to find a
rare, massive halo at high redshift, the corresponding abundance cannot be
estimated directly from our resimulations since these focus on ``special''
regions. It cannot be much lower than the $z=0$ abundance of galaxy clusters
like that initially selected, and in fact it turns out to be significantly
higher. Nevertheless, only an extremely small fraction of all cosmic mass is
locked up in objects as massive as the one we are tracking.

We illustrate these points in the lower two panels of Figure~\ref{fig:fig1}
using the Sheth \& Tormen (1999) formulae to estimate abundance and mass
fraction. Note that these estimates require extrapolation far beyond the
regime in which these formulae have been numerically tested; while undoubtedly
qualitatively correct, they may contain significant errors beyond $z\sim
10$. At $z\sim 50$ the comoving abundance of objects like the one we have
simulated is predicted to be similar to that of $10^{11}{\rm M_\odot}$ haloes
today, but such objects contain only a few times $10^{-7}$ of all cosmic
matter, corresponding approximately to $5.9\sigma$ peaks of the initial
gaussian density field. In contrast, the rich cluster we originally selected
corresponds only to a $2.7\sigma$ peak, while the $z=0$ descendent of our high
redshift halo corresponds to an even less rare $1.7\sigma$ peak. It is
interesting that a large excursion is visible at $z\sim 5$ in the plots of
Figure~\ref{fig:fig1}, which explains how the progenitor of the $10^{14}{\rm
M_\odot}$ object managed to ``beat out'' that of the more massive $z=0$
cluster we had originally chosen. In addition, there is a clear change in
behaviour in some of these plots after this redshift which is related to the
fact that the object is no longer subject to the condition that it must evolve
into an extreme system at later times (see also Figure~4 below).

\begin{figure*}
\vspace{-0.5cm}
\resizebox{10.0cm}{!}{\includegraphics{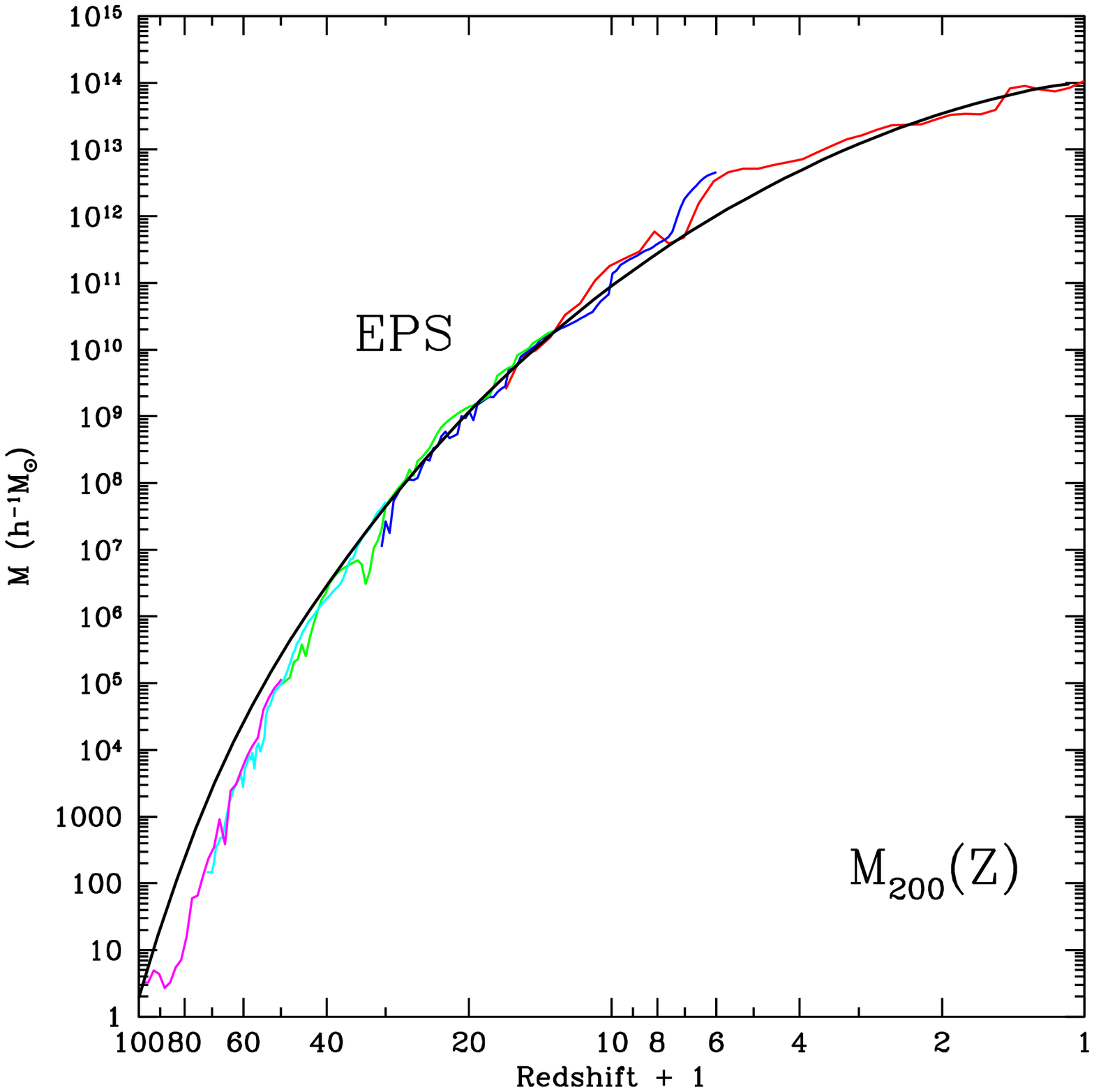}}
\end{figure*}

\begin{figure*}
\vspace{-0.9cm}
\hspace{-0.5cm}
\resizebox{9cm}{!}{\includegraphics{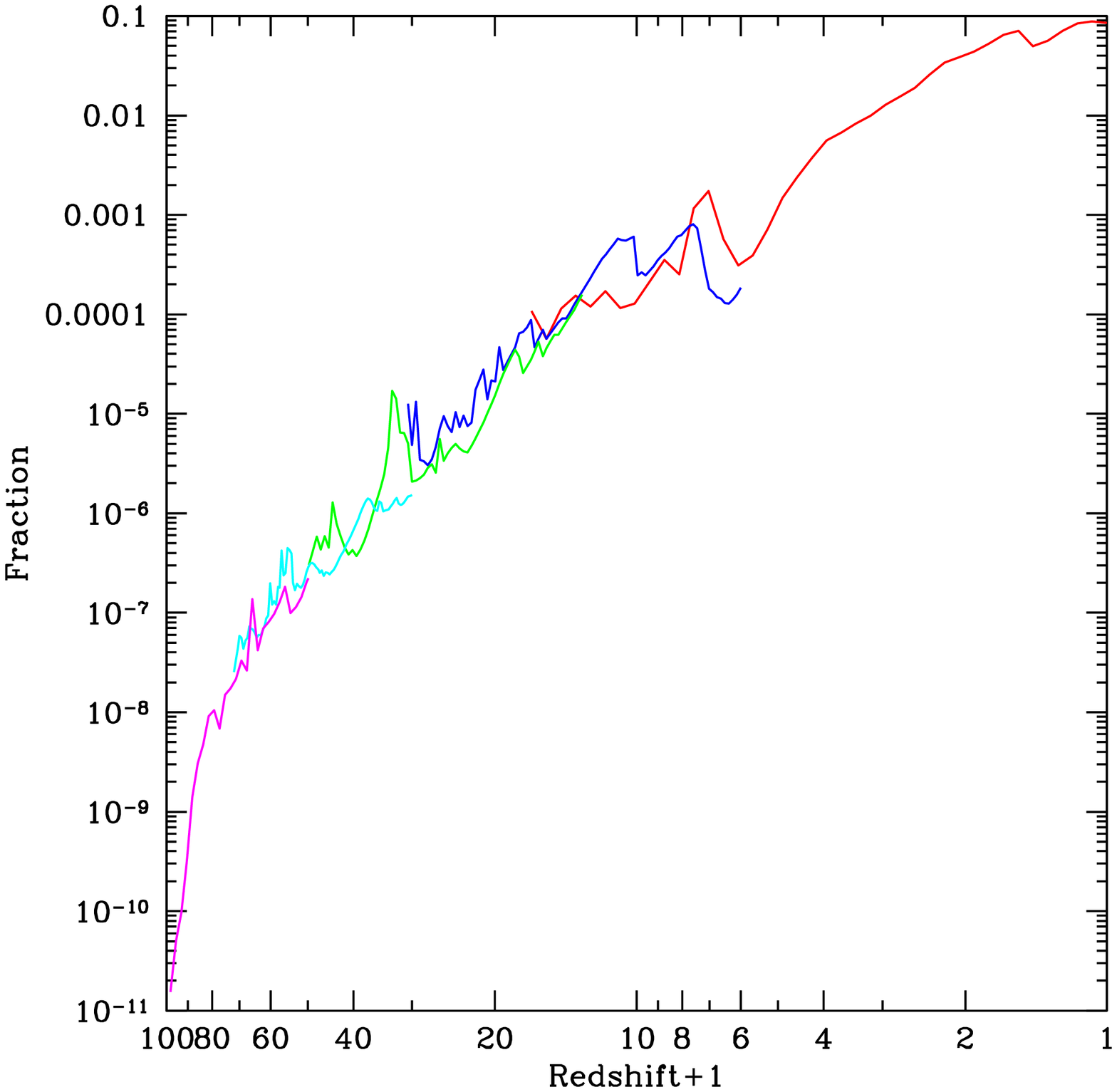}}%
\hspace{-0.1cm}\resizebox{9cm}{!}{\includegraphics{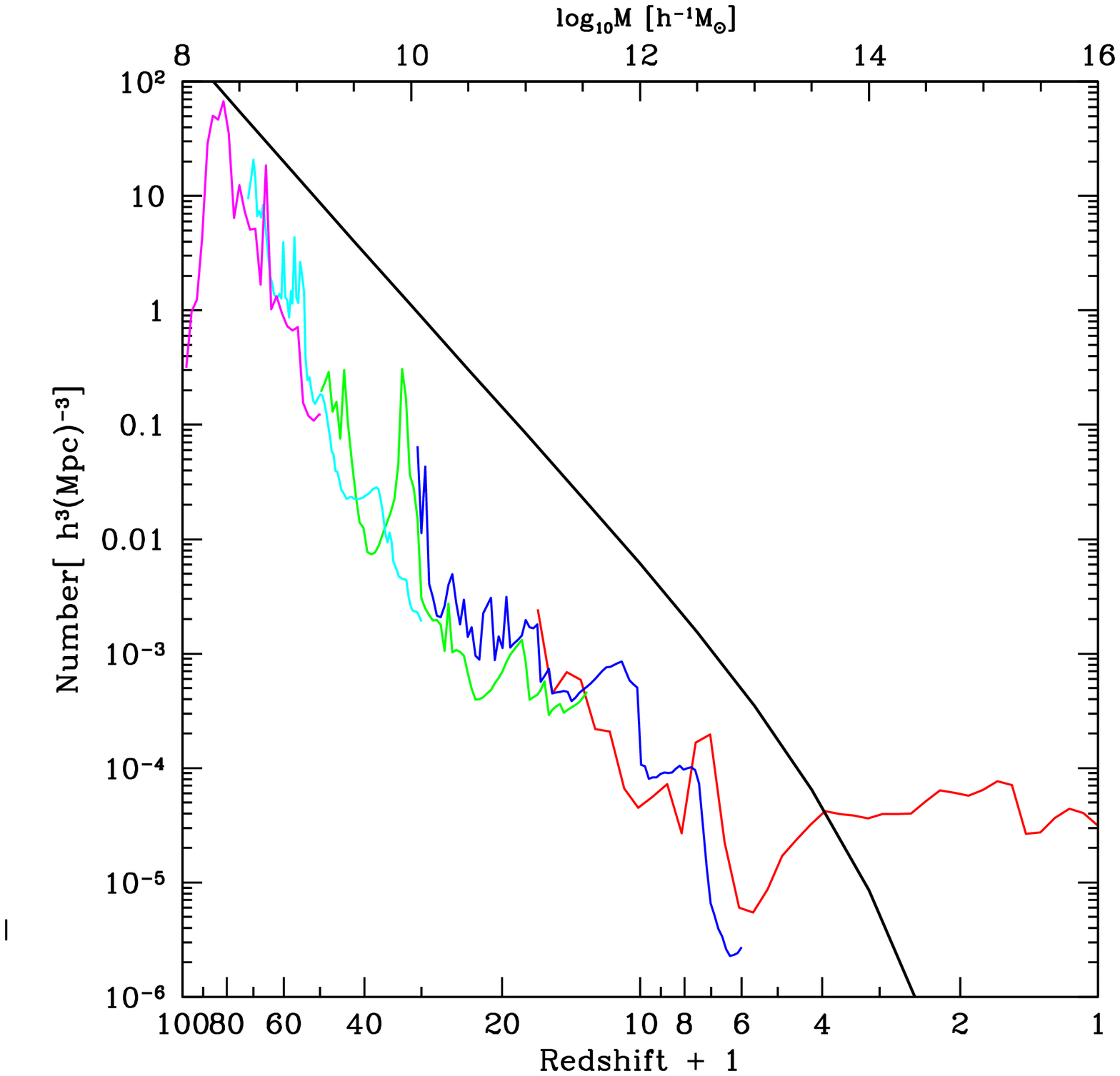}}\\%
\caption{(Top) The mass $M_{200}$ of the object we have followed since its
first progenitor was detected (at $z\sim 80$) until the present day. Different
line-types represent our different resimulations. The smooth curve is
predicted using extended Press-Schechter theory and is discussed in section
4.2. (Lower left) The fraction of all cosmic mass which at redshift $z$ is in
haloes with mass exceeding that plotted in the top panel.  (Lower right) The
comoving cosmic abundance of haloes at least as massive as our dominant
object. The results in the two lower panels are based on the formulae of Sheth
\&Tormen (1999). The thick solid line in the lower right panel is the
cumulative halo abundance at at $z=0$ predicted by these same formulae and
should be compared with the mass scale at the top of the plot.}
\label{fig:fig1}
\end{figure*}

Our R1 resimulation was carried out with the publicly available tree code
{\small GADGET-1.1} (Springel, Yoshida \& White 2001). Our other
resimulations used an improved {\small TREE-PM} code {\small GADGET-2}
(Springel 2005).

\subsection{Halo identification}
Two of the most common methods for identifying haloes in $N$-body simulations
are the friends-of-friends (FOF) algorithm of Davis et al. (1985), and the
spherical overdensity (SO) algorithm described by Lacey \& Cole (1994).
Advantages of FOF are that it does not impose any fixed shape on the haloes,
and that it is fast. However, it occasionally links separate haloes over a
chance bridge of particles, and the most massive haloes in a large simulation
are often such multiple objects.  In the limit of large particle number, FOF
haloes are approximately bounded by an isodensity contour.

In the SO algorithm, the mass of a halo is evaluated in a spherical
region. There is only one free parameter, the mean overdensity $\delta$ within
this region. There are, however, many possible ways to determine its
centre. For practical purposes, most of these are equivalent. In our
implementation, the centre is determined iteratively. A local maximum of the
density field, estimated by the standard {\small SPH} method, is taken as an
initial guess. A sphere is then grown outward until it reaches the desired mean
overdensity. The centre-of-mass of this sphere is taken as the next guess at
the centre and the procedure is repeated.  After several iterations, the
motion of the centre becomes small.  The SO algorithm rarely identifies haloes
with multiple major concentrations, but it is slower than the FOF algorithm
and it does impose a fixed spherical shape when determining halo masses

\begin{figure}
\centerline{\psfig{figure=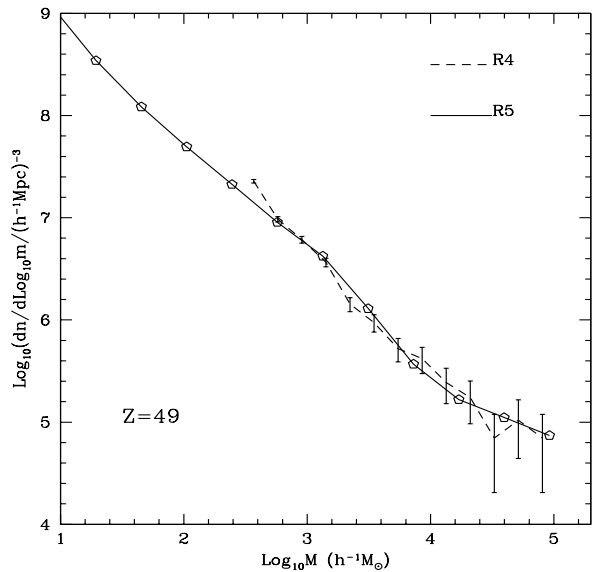,width=250pt,height=250pt}}
\caption{Halo mass functions at $z=49$ for the common region of the~R4 and~R5
simulations. Haloes were identified and their masses estimated using the
SO(180) algorithm. Error bars show Poisson uncertainties in the counts for the
R4 haloes.}
\label{fig:fig2}
\end{figure}

As we shall see in later sections, the first massive objects are found in
overdense regions containing many smaller haloes which line up along filaments
and sheets. In this situation we find that the FOF halo identification
algorithm is quite dependent on the mass resolution of the simulation.  For
example, using a linking length of 0.2 times the mean interparticle separation
(at the cosmic mean density), the halo mass functions in corresponding regions
at $z=49$ of the R4 and R5 resimulations are very different; $19$ per cent of
all high resolution particles are identified as a single halo in R5, but the
most massive halo is much smaller in R4. On the other hand, as can be seen in
Fig.~\ref{fig:fig2}, the abundance of haloes selected by SO assuming
$\delta=180$ (again relative to the cosmic mean) agrees well in the two
resimulations down to the resolution limit of R4. In the following we use the
SO(180) algorithm to identify all haloes, unless otherwise stated, although
we often quote masses and radii as $M_{200}$ or $r_{200}$ referring to a
spherical region with mean density 200 times the {\it critical} value.

\section{The evolution of extreme haloes}

In this section we study the growth of the main halo in our simulation
sequence and consider how its internal structure changes with time. We wish to
test whether the halo and its environment at high redshift are similar to a
suitably scaled version of low redshift structures.  It has often been argued
that dark matter evolution should be effectively self-similar in hierarchical
clustering. This could only be strictly true in an Einstein-de Sitter universe
with a power-law initial fluctuation spectrum (e.g. Efstathiou et al. 1988;
Smith et al.  2003). Some deviations are expected in the concordance
$\Lambda$CDM cosmology, both because of the influence of the cosmological
constant at low redshift, and because of the curvature of the initial power
spectrum. In addition, the fact that the power spectrum slope approaches the
critical value $-3$ at high wavenumber implies a breakdown at early times of
the critical assumption underlying the hierarchical clustering model, namely
that small things form first. Our sequence of resimulations provides an
opportunity to explore whether these effects lead to qualitatively different
behaviour at early times. We begin by displaying images of the main halo
and its immediate surroundings at the end of each of our resimulations.
These are the times when the structure is best resolved. Subsequent
subsections consider the evolution in characteristic temperature, in density
profile and in substructure content of the principal halo.

\subsection{Images}

\begin{figure*}
\resizebox{8cm}{!}{\includegraphics{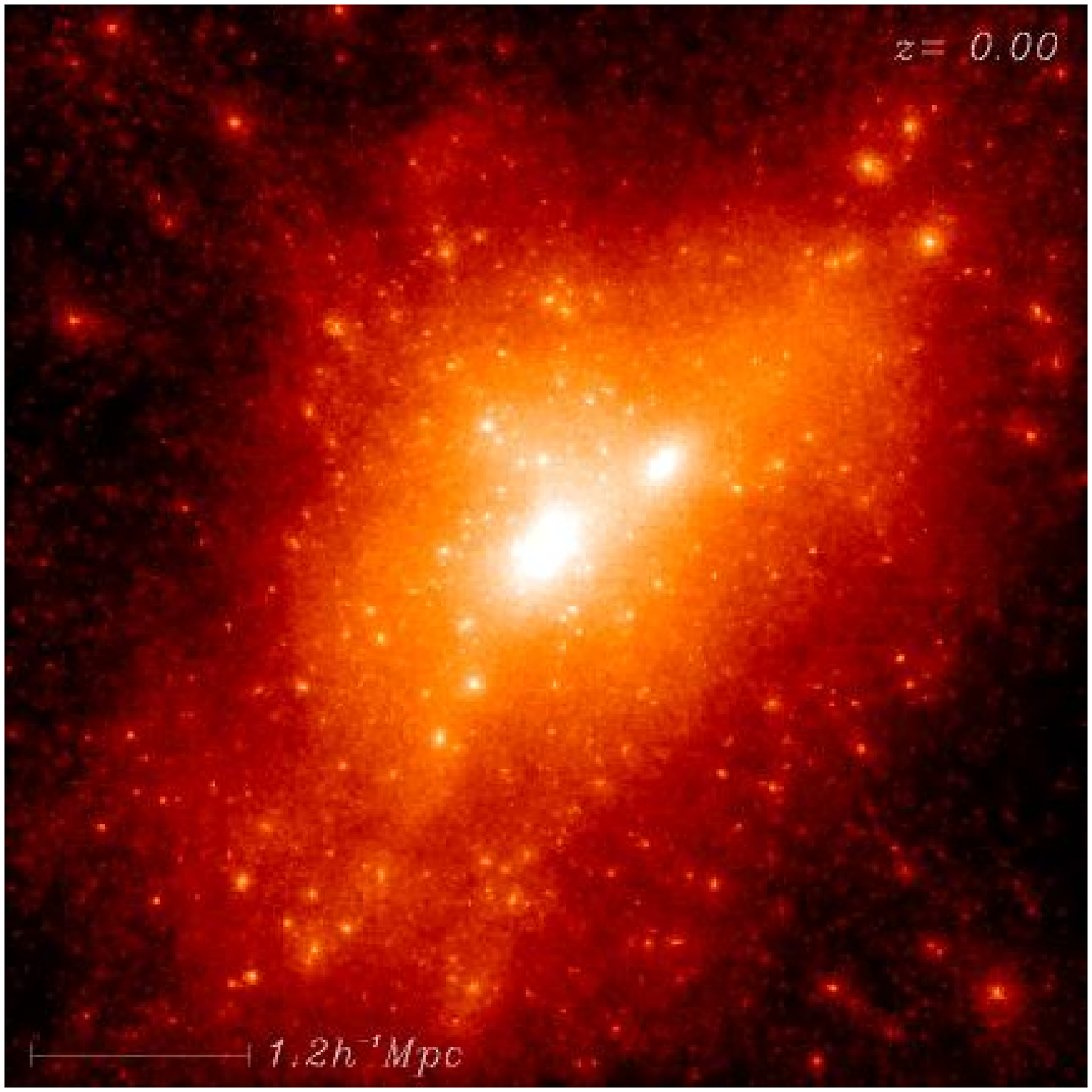}}%
\hspace{0.13cm}\resizebox{8cm}{!}{\includegraphics{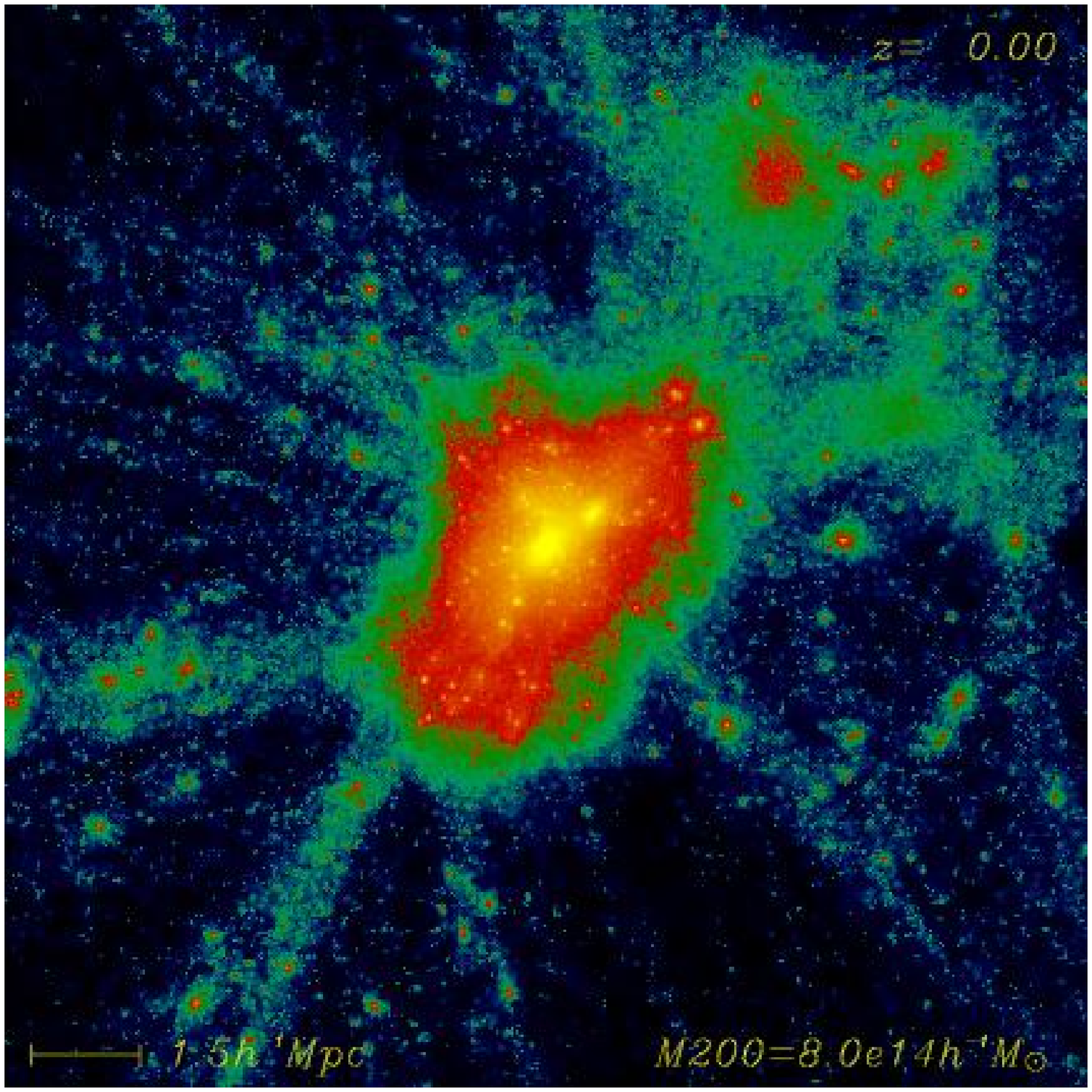}}\\%
\resizebox{8cm}{!}{\includegraphics{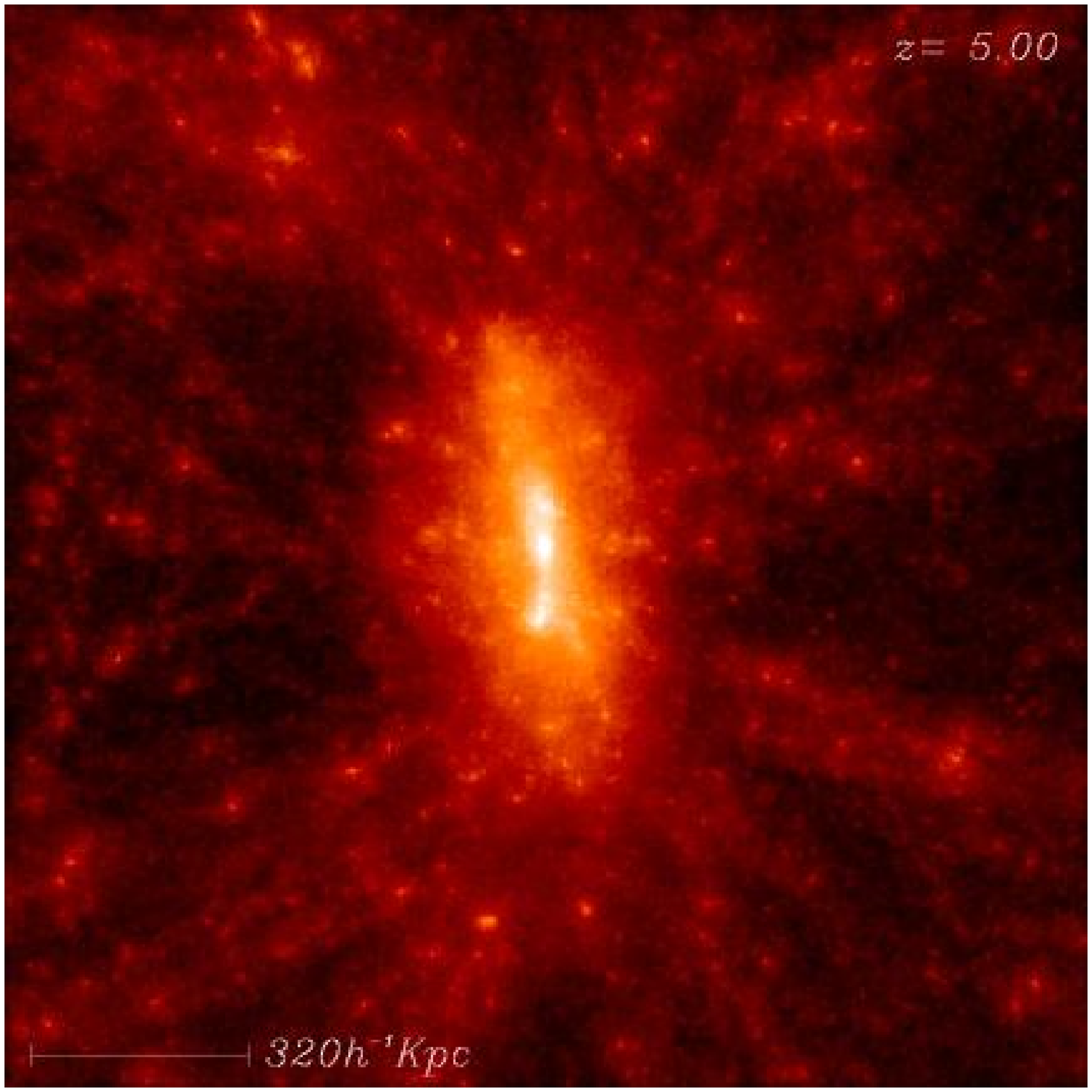}}%
\hspace{0.13cm}\resizebox{8cm}{!}{\includegraphics{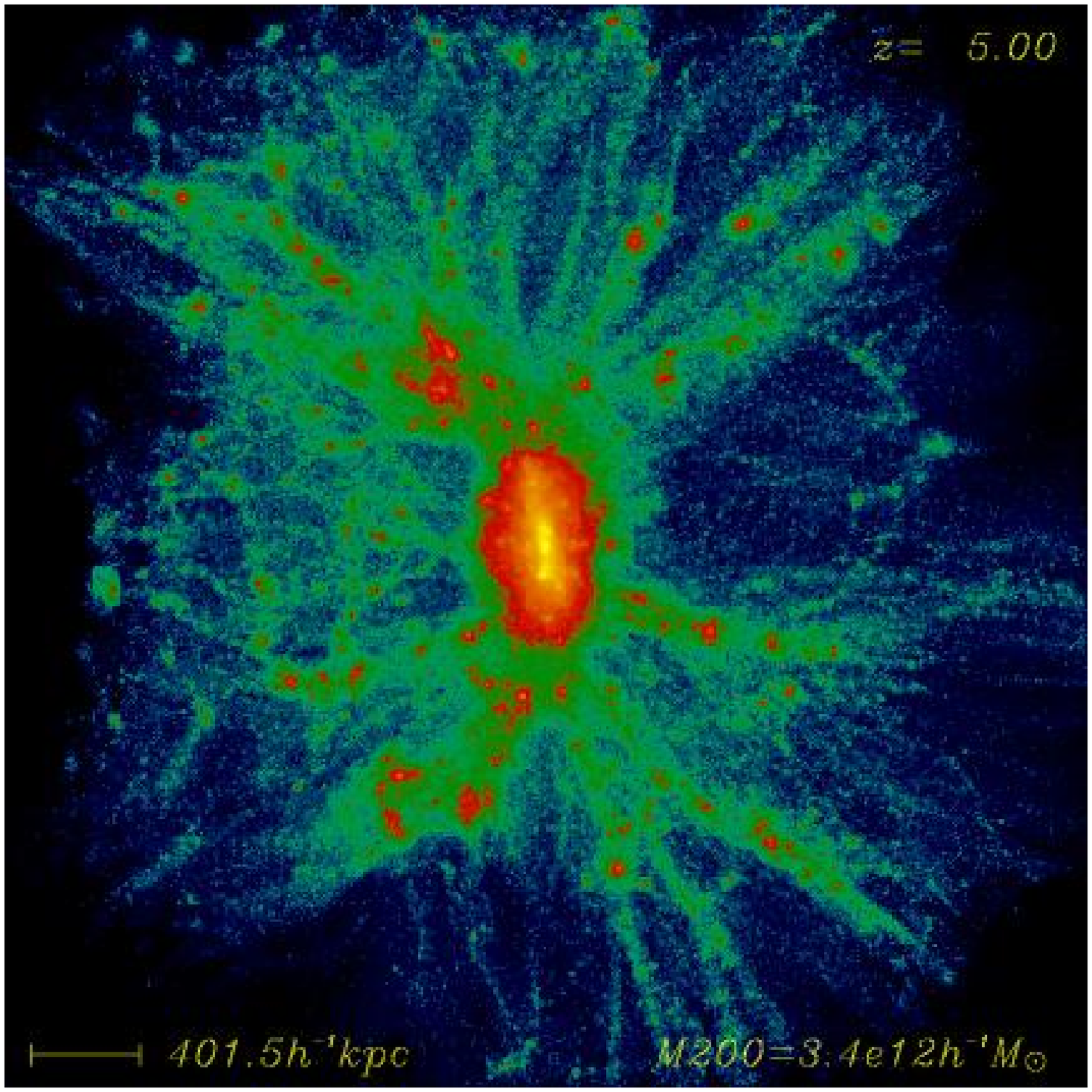}}\\%
\resizebox{8cm}{!}{\includegraphics{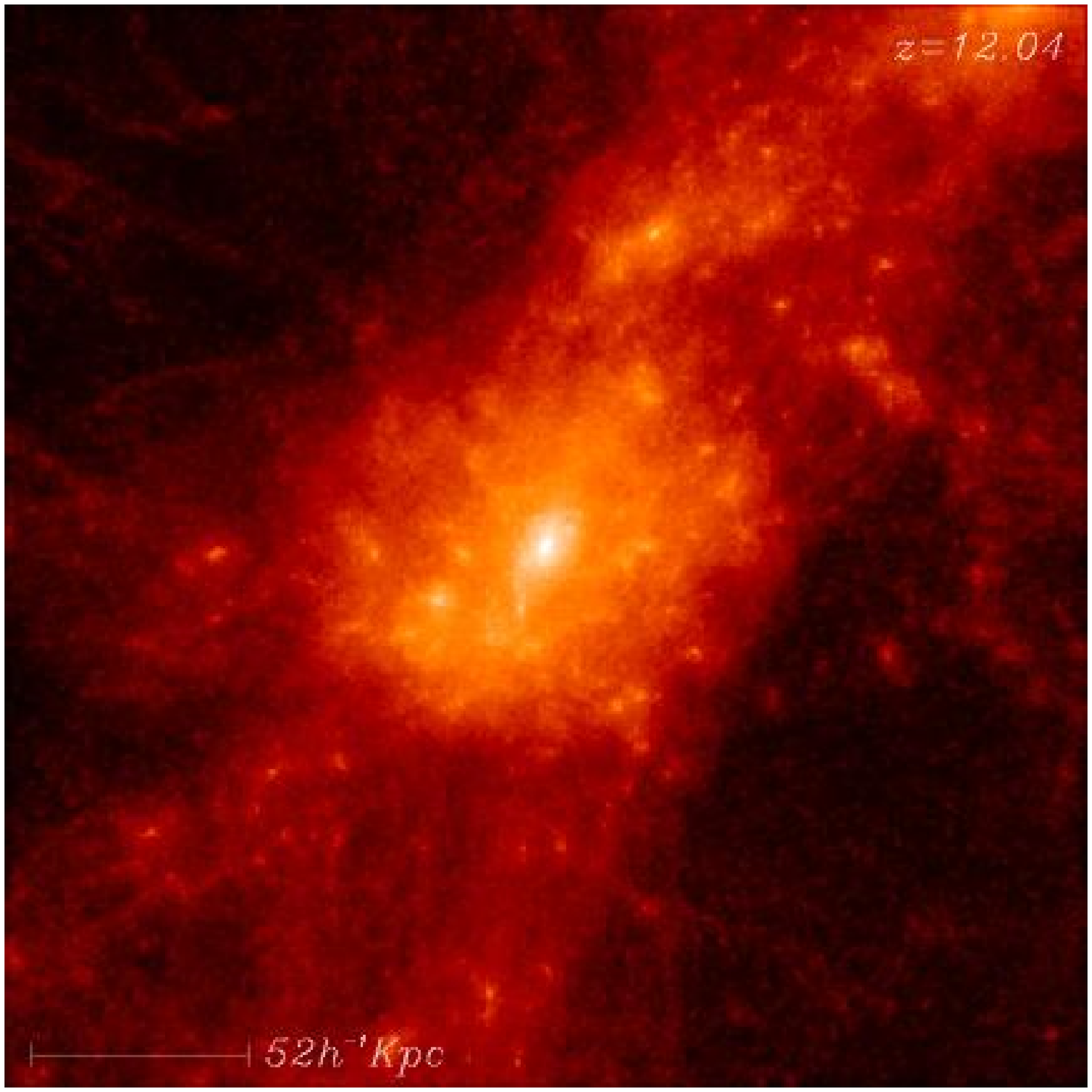}}%
\hspace{0.13cm}\resizebox{8cm}{!}{\includegraphics{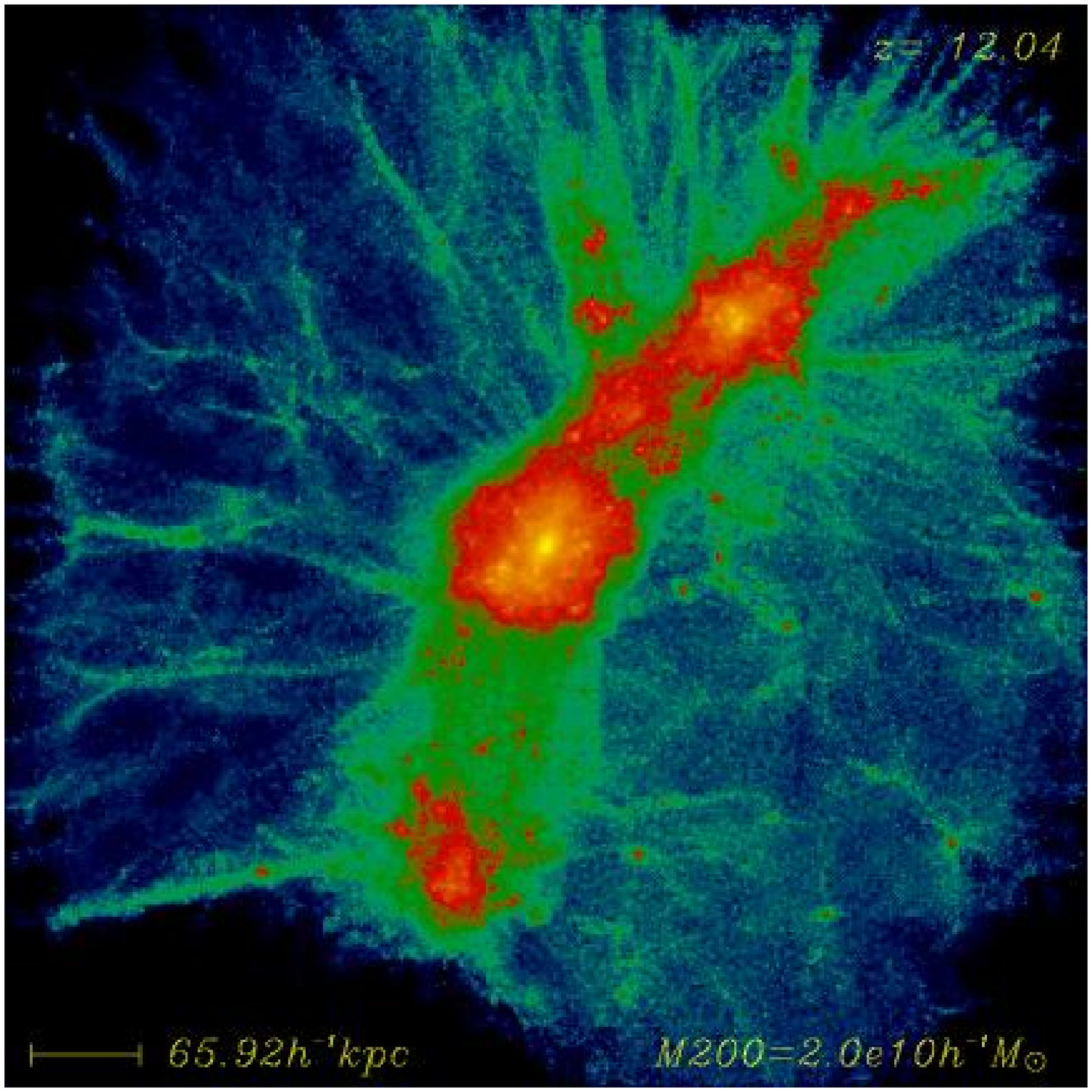}}%
\end{figure*}

\begin{figure*}
\resizebox{8cm}{!}{\includegraphics{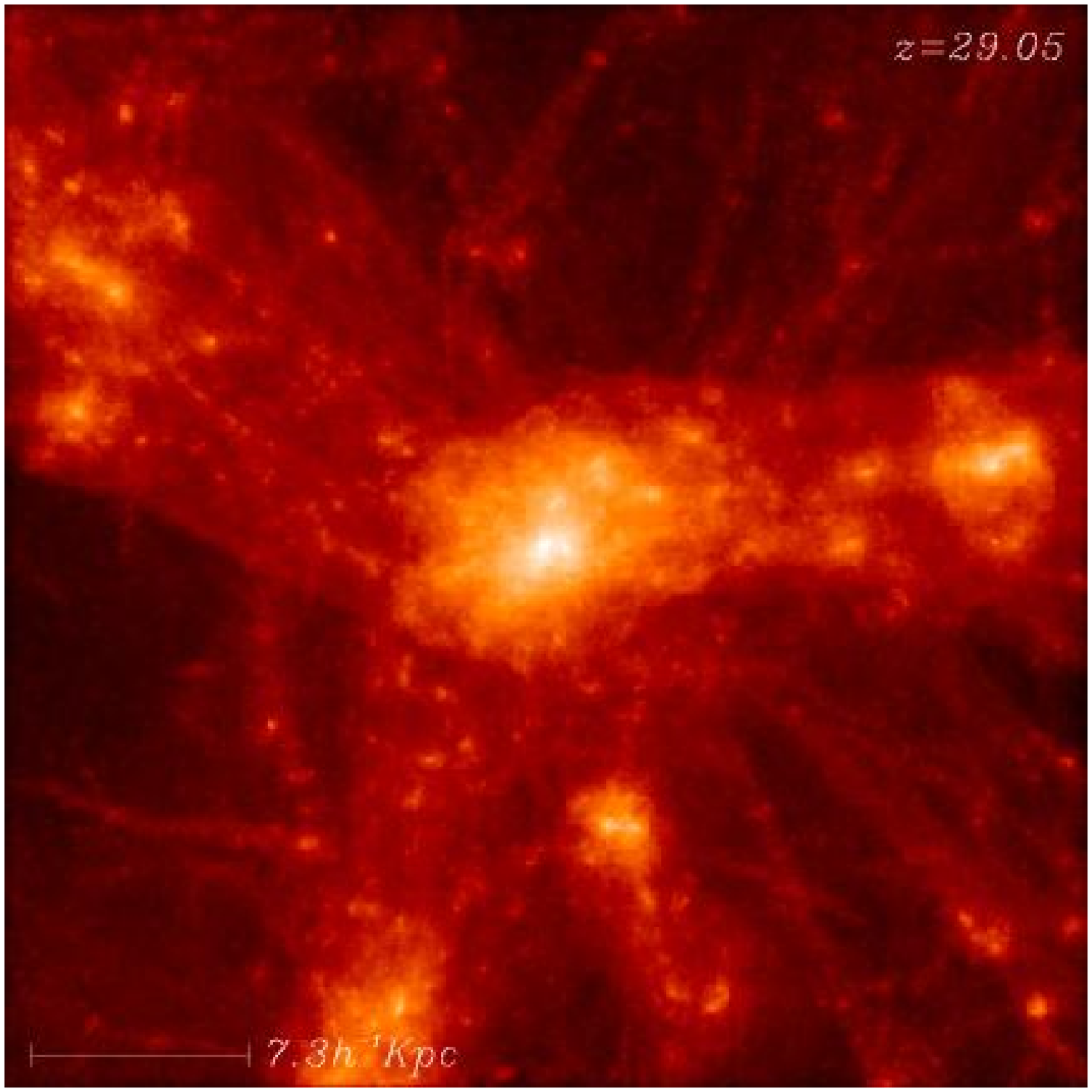}}%
\hspace{0.13cm}\resizebox{8cm}{!}{\includegraphics{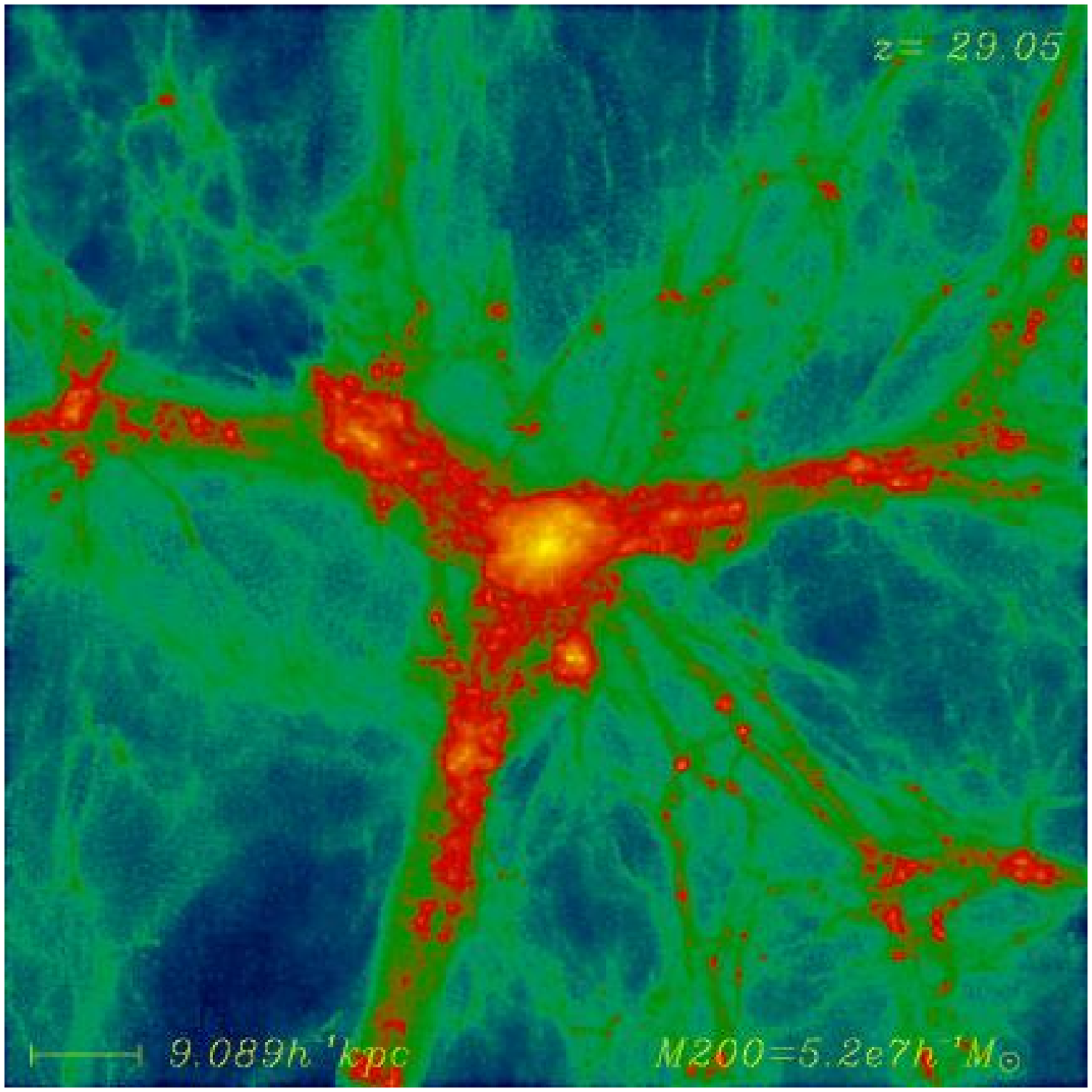}}\\%
\resizebox{8cm}{!}{\includegraphics{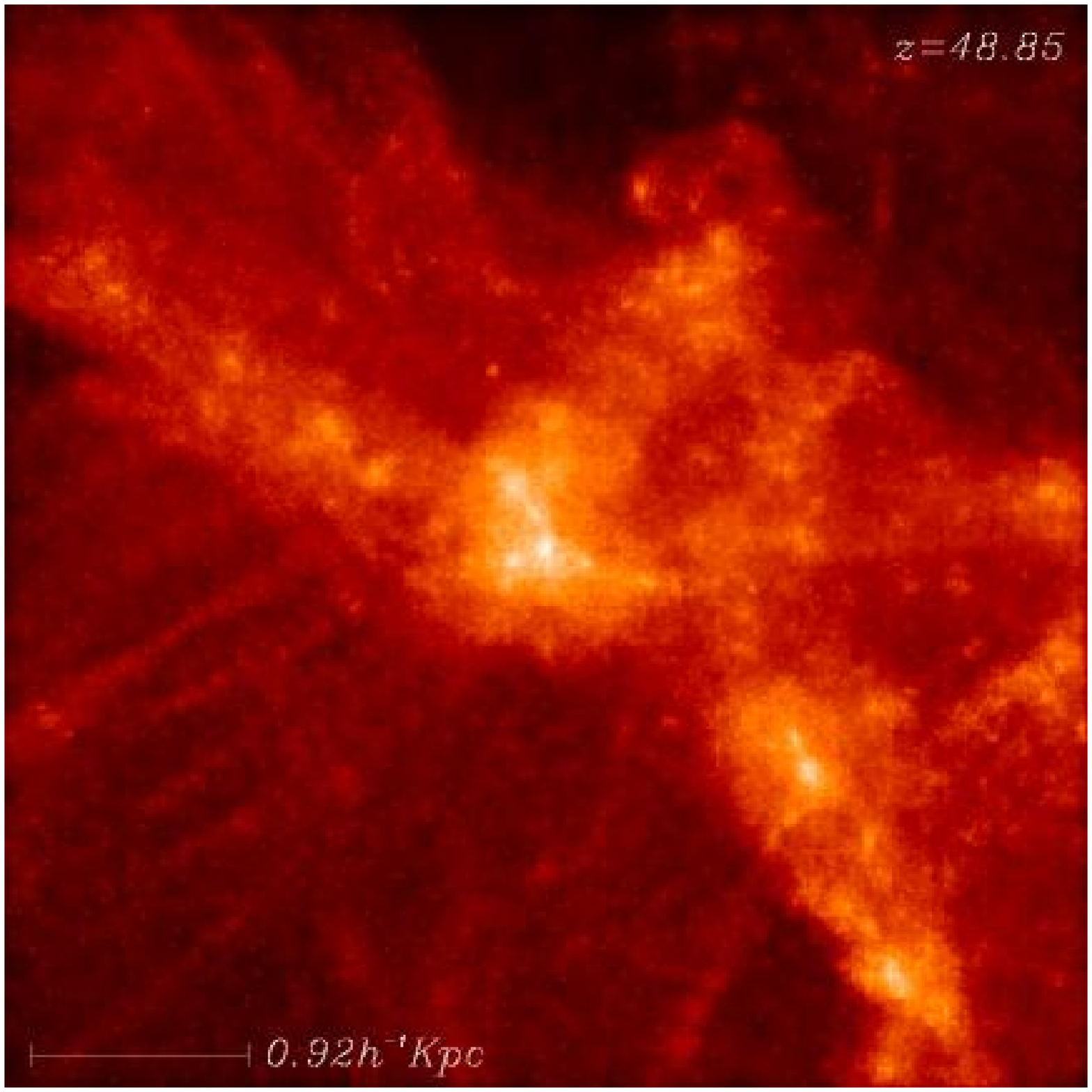}}%
\hspace{0.13cm}\resizebox{8cm}{!}{\includegraphics{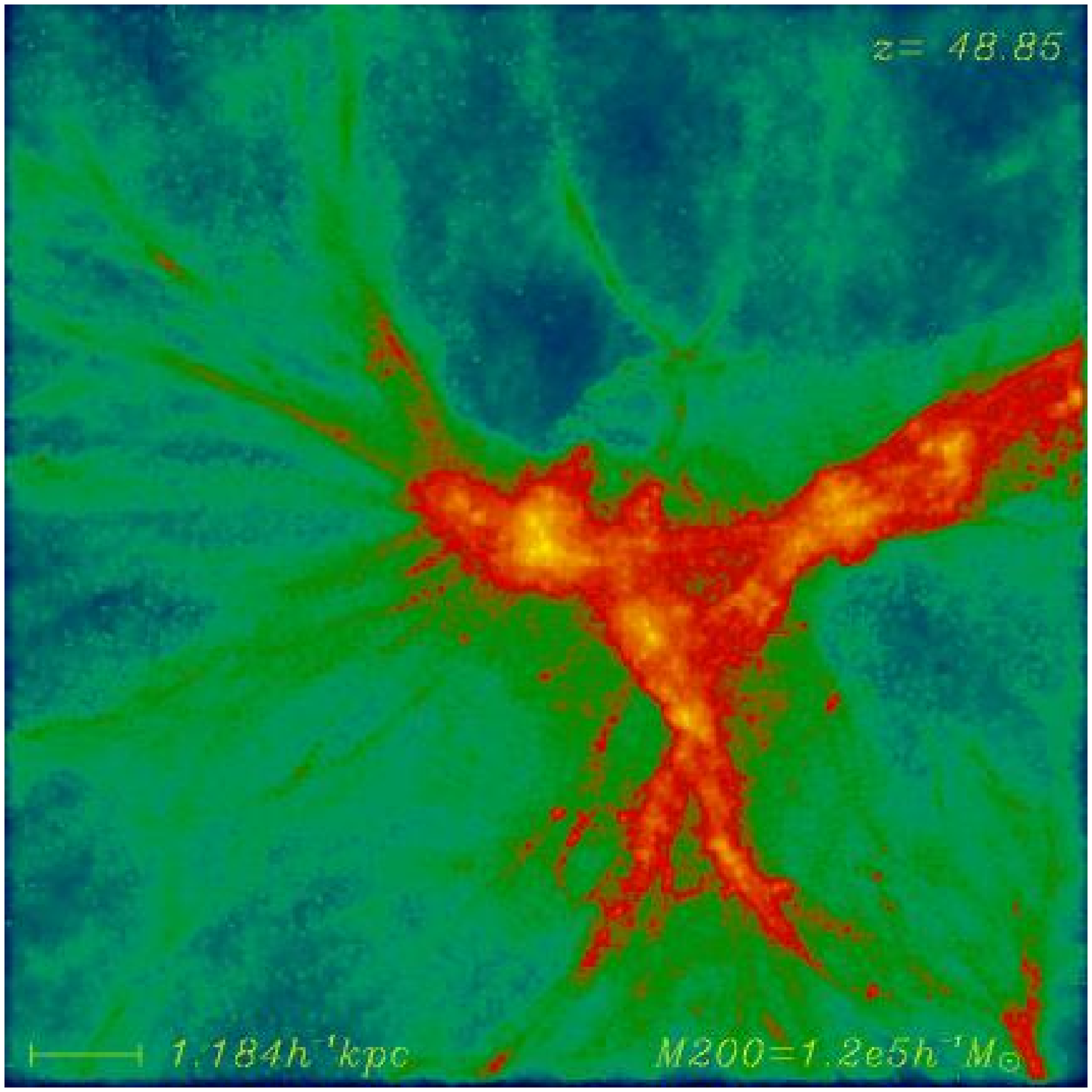}}\\
\vspace{0.1cm}
\hspace{-8cm}\resizebox{8cm}{!}{\includegraphics{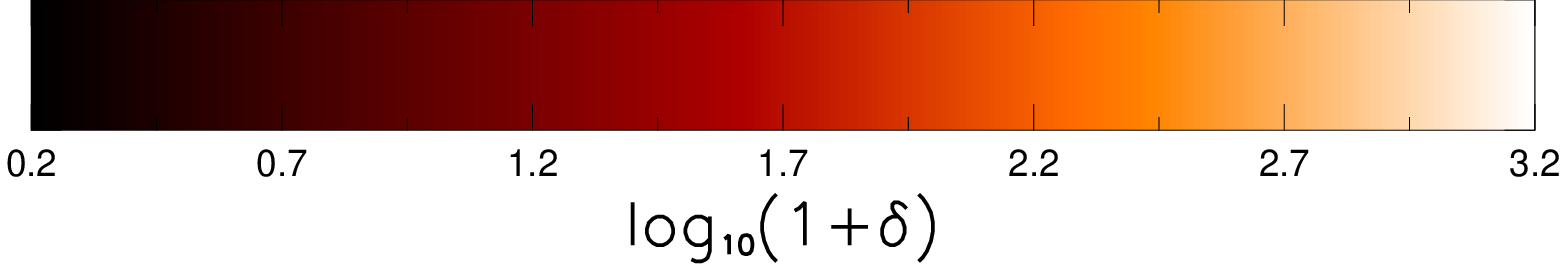}}\\
\vspace{-1.35cm}
\hspace{8.13cm}\resizebox{8cm}{!}{\includegraphics{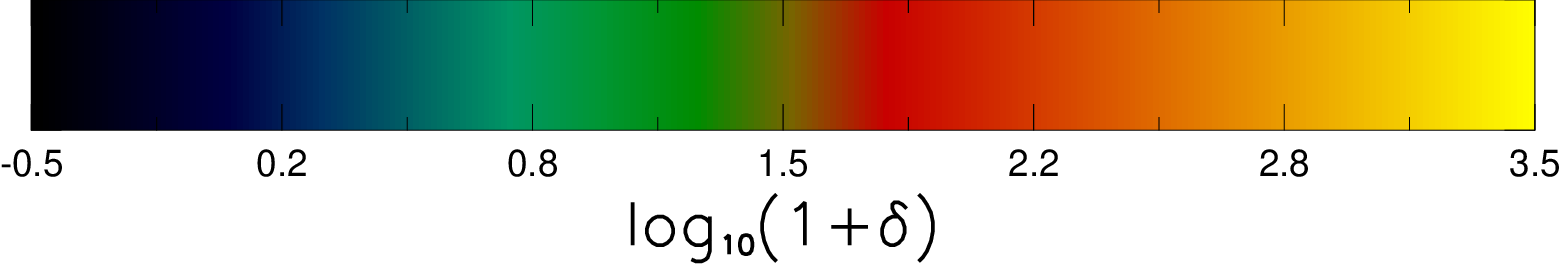}}\\

\caption{Projected dark matter density in and around the most massive haloes
present at the final time of each of our resimulations.  The left-hand panels
show the dark matter distributions within cubes of side $4r_{200}$ centred on
the main halo. The right-hand panels show distributions in slices $10r_{200}$
on a side and $2r_{200}$ thick, also centred on the main halo. In each image,
the density field is normalized to the projected mean cosmic density and
depicted on a logarithmic scale. The colour scales are identical at the
different times in each series of images but differ between the left and right
series.}
\hspace*{1.36cm} \\
\label{fig:fig3}
\end{figure*}

The distribution of dark matter in and around the most massive halo at the end
of each of our resimulations is illustrated in Fig.~\ref{fig:fig3}. Note
again that the $z=0$ halo is not the descendent of the haloes shown at earlier
times. The left-hand column depicts one projection of the density in a cube of
side $4\times r_{200}$ centred on the main halo. (In comoving units, $r_{200}$
ranges from $1.5$~Mpc for the R1 halo at $z=0$ to $1.2$~kpc for the R5 halo at
redshift $z=49$; see Table 1.)  The surface density is normalized to the
cosmic mean and the color table represents the dimensionless surface density,
$1+\delta$, on a logarithmic scale. In all cases, a centrally concentrated
object is clearly visible at the centre of the image, but as the redshift
increases the matter distribution becomes increasingly anistropic. At $z=12$
and earlier very strong filaments surround the central halo and appear less
fragmented into individual smaller objects than at later times.

The right-hand column of Fig.~\ref{fig:fig3} illustrates the larger scale
environment of these main haloes. The images here have side $10r_{200}$, are
projections of a region of depth $2r_{200}$, and again all represent the
projected overdensity using an identical colour scale.  The qualitative change
in morphology with redshift is more dramatic in these plots. As we go back in
time the main halo becomes less dominant and less regular, the ``filamentary''
structures become heavier and smoother, and substructure both within and
outside the main object becomes less evident. At the earliest times, the
filaments extend over the full region plotted and are the most striking
structure within it. Note that with the exception of the latest pair ($z=5$
and $z=0$) all the matter shown in each early image is contained within the
central object of the next later one.

\subsection{The growth in halo temperature}
\begin{figure}
\centerline{\psfig{figure=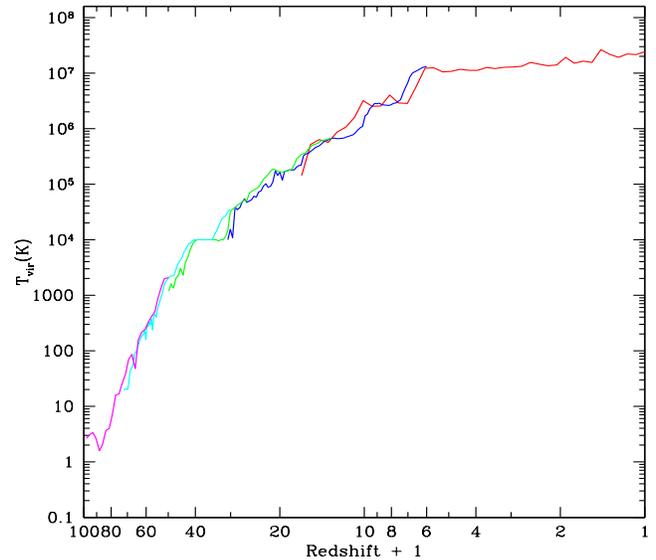,width=250pt,height=250pt}}
\caption{Increase with time of the characteristic temperature (estimated
from the maximum circular velocity) of the principal halo in our sequence of
resimulations.} \label{fig:fig4}
\end{figure}

The rapid growth in mass of our principal halo was already shown
in the top panel of Fig.~\ref{fig:fig1}. At each time we can
estimate a representative temperature for the gas associated with
this halo from its maximumum circular velocity, $V_c$:
\begin {equation}
T= \mu m_pV_c^2/(2k_B)\,
\end {equation}
where $\mu m_p$ is the mean molecular weight of the gas and $k_B$ is
Boltzmann's constant. This ``virial'' temperature is plotted as a function of
redshift in Fig.~\ref{fig:fig4}. We assume $\mu=1.22$ from the earliest
redshift until the temperature has risen to $10000$K.  At this time we assume
that the gas is collisionally ionised and has $\mu=0.59$ thereafter; during
the transition the halo temperature is taken to be constant at ${\rm T}_{\rm
Vir}=10000$K. Like the mass, the halo temperature increases very rapidly at
early times. Already by $z=49$, it has reached $\sim 2000$K when, for the high
densities predicted at these redshifts, molecular hydrogen should form
sufficiently rapidly to provide efficient radiative cooling. As many recent
simulations have shown, this is likely to lead to the formation of a massive
star at the centre of our object (Abel et al 1998, 2002; Bromm et al. 2002;
Yoshida et al. 2003). The critical temperature for collisional ionisation
$T\sim 10^4$K is reached by $z=39$ and the greatly increased efficiency of
cooling at this point could lead to a starburst and the formation of a
minigalaxy. Whether this actually happens depends on the evolution of the
first star.  Its ionising radiation and the strong shock it produces if it
goes supernova could already have affected surrounding gas over a large region
by $z=39$ (e.g.  Yoshida, Bromm \&Hernquist 2004). We study these issues as
well as others associated with the formation of the first stars in a companion
paper (Reed et al. 2005c).

\subsection{The evolution of halo density profiles}
Simulations of structure formation in hierarchical clustering cosmogonies
(including the concordance $\Lambda$CDM model) produce dark matter haloes with
density profiles which can be fit by the simple formula proposed by Navarro,
Frenk \& White (1996, 1997; NFW),
\begin{equation}
\rho(r)={\rho_s \over (r/r_s) (1+(r/r_s)^{2}}.
\label{eq:nfw}
\end{equation}
Here $r_s$ is a characteristic radius where the profile declines
locally as $r^{-2}$, and $\rho_s/4$ is the density at $r_s$. At
the relatively low redshifts where this formula has been tested,
there is a strong correlation between $r_s$ and $\rho_s$ which
depends on redshift, on the global cosmological parameters, and on
the initial power spectrum. The larger the mass of a halo, the
lower its characteristic density, reflecting the lower density of
the universe at the (later) time when more massive systems were
typically assembled.  The applicability of this formula in the
innermost regions of haloes has been controversial, but recent
work suggests that it is a reasonable fit to most haloes over the
radial range $0.01r_{200}<r<r_{200}$; the behaviour at yet smaller
radii remains a topic of debate (Navarro et al. 2004; Reed et al.
2005a; Diemand et al. 2004).

\begin{figure*}
\resizebox{8cm}{!}{\includegraphics{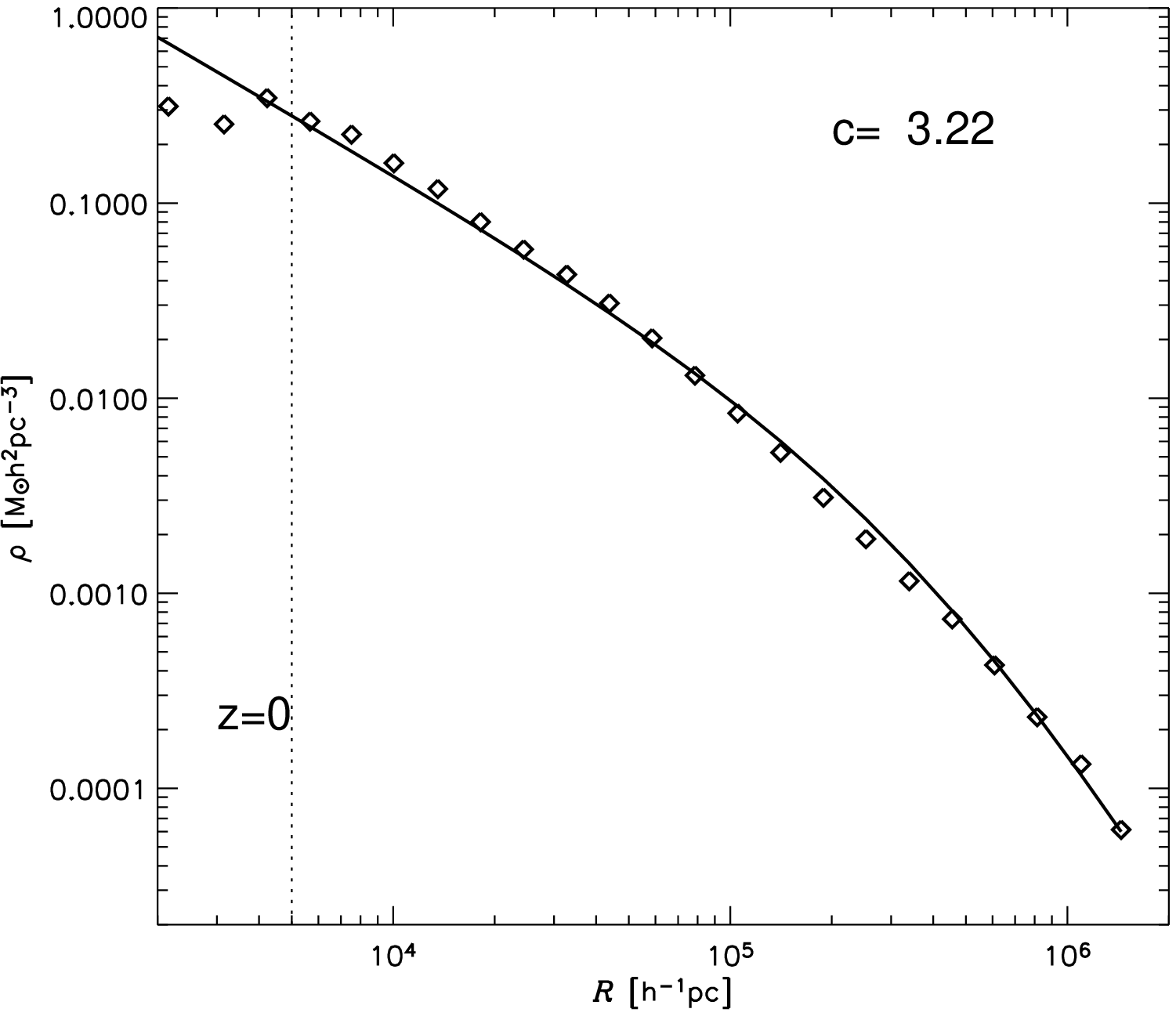}}%
\hspace{0.13cm}\resizebox{8cm}{!}{\includegraphics{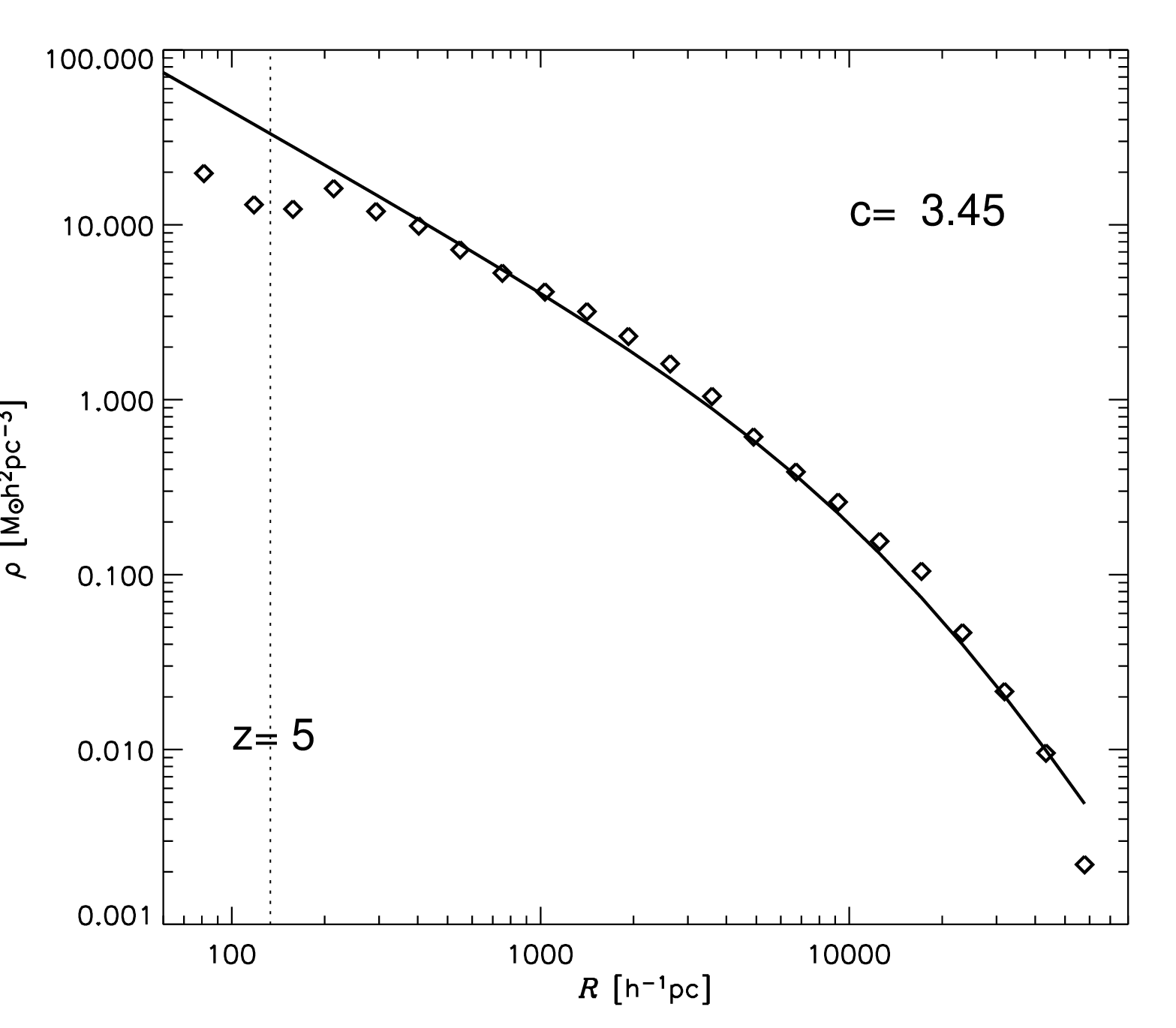}}\\%
\resizebox{8cm}{!}{\includegraphics{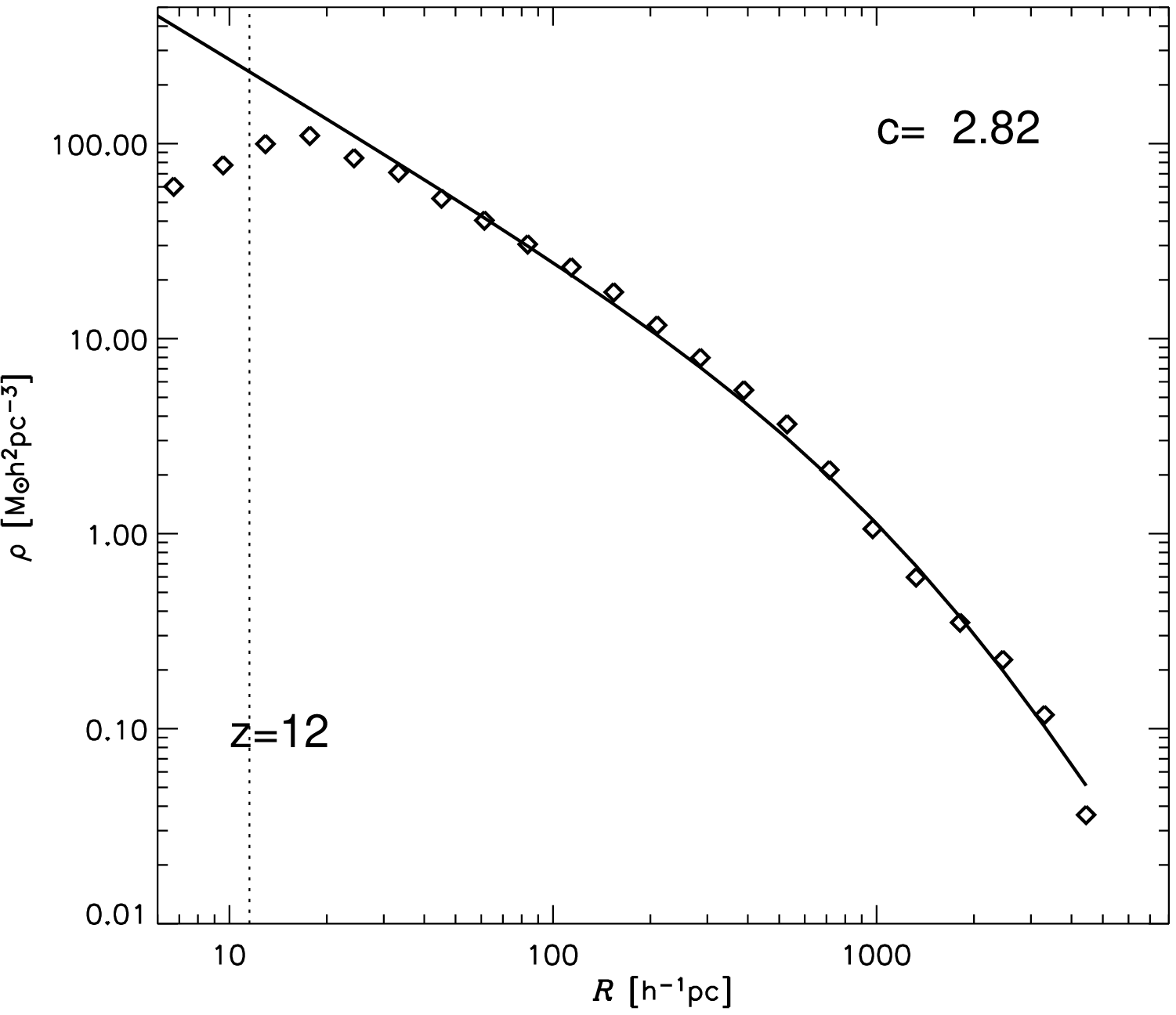}}%
\hspace{0.13cm}\resizebox{8cm}{!}{\includegraphics{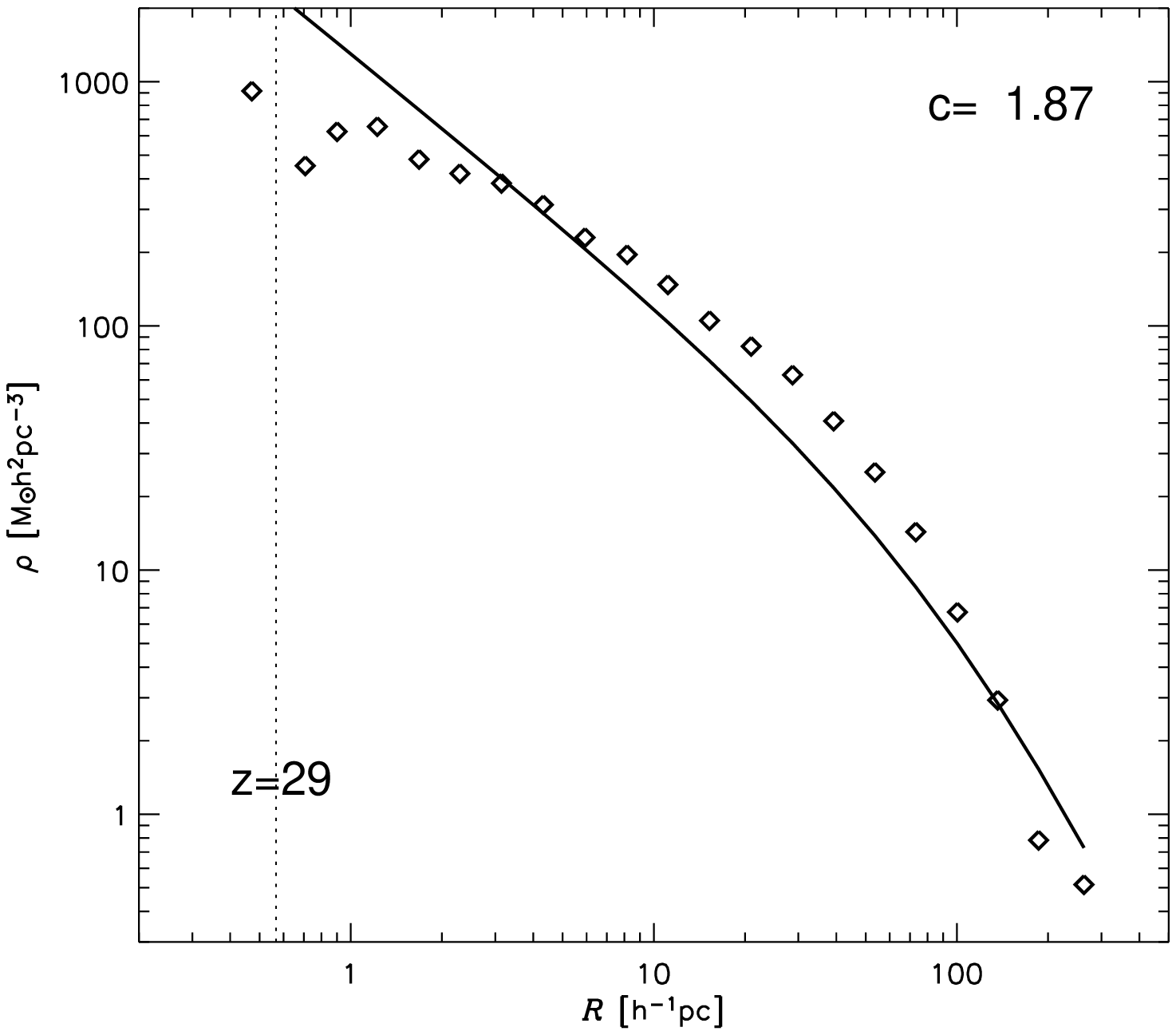}}\\%
\resizebox{8cm}{!}{\includegraphics{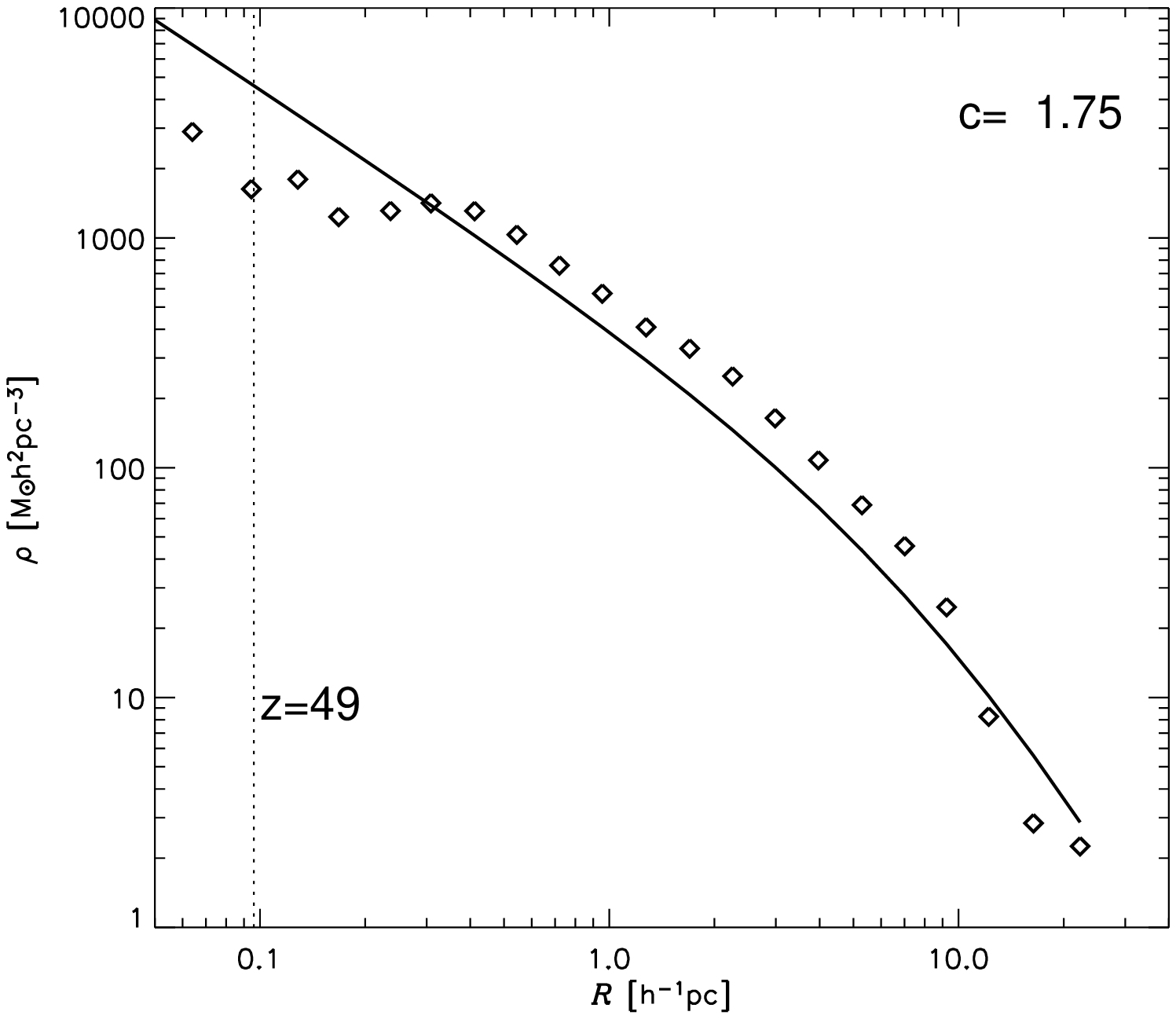}}%
\hspace{0.13cm}\resizebox{8cm}{!}{\includegraphics{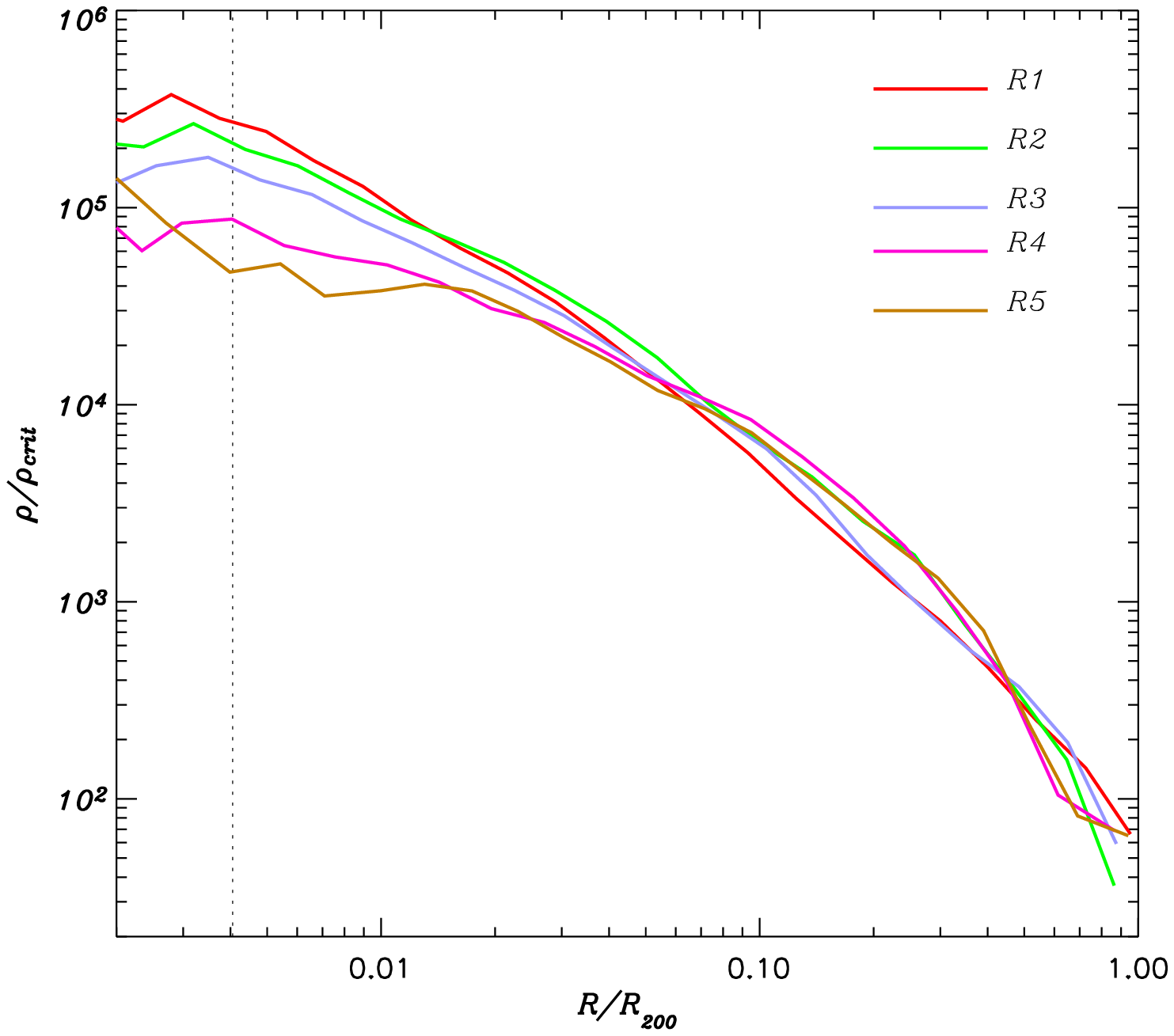}}%
\caption{Density profiles for the final halo in each resimulation.  Open
squares show the mean density within spherical shells plotted against shell
radius. Both axes are given in physical (rather than comoving) units. The
solid lines are fits to an NFW profile and weight each point outside the
gravitational softening radius equally. They imply the concentration
parameters $c=r_{200}/r_s$ given in each panel. Dashed vertical lines show the
gravitational softening in each resimulation.  The five profiles are
superposed in the bottom right plot after scaling radii to $r_{200}$ and
densities to the critical density. This emphasises the evolution of shape with
redshift. } \label{fig:fig5}
\end{figure*}

Roughly two million particles lie within the virial radius of the final halo
in all our resimulations except R5, whose halo has only $\sim 220000$
particles within $r_{200}$. The density profiles are very well sampled and
this allows us to investigate the internal structure of these extreme objects
as a function of redshift. In Fig.~\ref{fig:fig5} we show the density
profile of the most massive halo in each resimulation at the final time,
together with the best-fit NFW profile. The NFW formula is a good fit to the
data over the full range plotted at the three later times. At $z=29$ and
$z=49$, however, the fits are quite poor; the profiles are shallower than the
analytic formula at small radii and steeper at large radii.

In the lower right-hand panel of the figure, we superpose the density profiles
so that their shapes can be compared. When scaled by $r_{200}$, the profiles
vary systematically with redshift, their inner slope and concentration
decreasing with increasing redshift. Note that the profiles in R1 and R2 are
quite similar, as are the profiles in R4 and R5. The profile in R3 is
intermediate in behaviour. The R4 and R5 profiles look qualitatively different
from the others, reinforcing the impression from the images in
Figure~~\ref{fig:fig3}. These changes presumably reflect the very steep
spectrum of the $\Lambda$CDM cosmology and the resultant very rapid growth 
of structure on these small mass scales.

\subsection{Substructure in early haloes}

The high resolution achieved by N-body simulations in recent years has enabled
detailed study of the substructure expected within dark haloes in the standard
cosmogony (Ghigna et al. 2000; De Lucia et al. 2004; Diemand et
al. 2004b,Gao et al. 2004a,b). Subhaloes are the surviving cores of
objects which fell together during the hierarchical assembly of the
halo, and it is natural to attempt to relate their number and
properties to observed substructures such as individual galaxies
within galaxy clusters (Springel et al. 2001b) or satellite galaxies
within the haloes of Milky Way-like systems (Klypin et al. 1999; Moore et
al. 1999). Unfortunately, both observational and theoretical
arguments show the correspondence between subhaloes and galaxies to be
quite complicated and to depend strongly on the history of the
subhaloes rather than just on their mass and position (Springel et
al. 2001b; Stoehr et al.  2002; Gao et al. 2004a,b; Nagai \&~Kravtsov,
2005).

\begin{figure}
\centerline{\psfig{figure=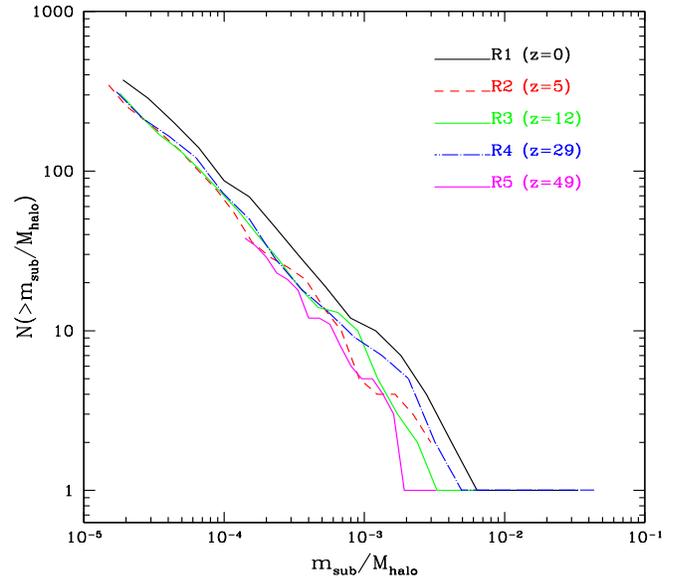,width=250pt,height=250pt}}
\caption{Cumulative subhalo mass functions for the most massive halo at the
final time in each of our sequence of resimulations.} \label{fig:fig6}
\end{figure}

Because of the extremely rapid build-up of the main object in our resimulation
sequence and the relatively smooth appearance of the structure that surrounds
it at early times, it might seem {\it a priori} likely that substructure
should be less prominent in early haloes than in low redshift
systems. Figure~\ref{fig:fig6} demonstrates that this is {\it not} the
case. We plot the cumulative subhalo mass function for the most massive halo
at the final time in each of our resimulations. Masses are normalised to the
value of $M_{200}$ for the massive halo and only subhaloes within $r_{200}$
are included. The subhaloes here were identified using the {\small SUBFIND}
algorithm described in Springel et al (2001b). Remarkably, the substructure
mass function is very similar at all redshifts except $z=0$.  Subhaloes in the
massive cluster halo in R1 are roughly a factor of 1.5 more abundant at given
mass fraction than at earlier times.

\begin{figure}
\centerline{\psfig{figure=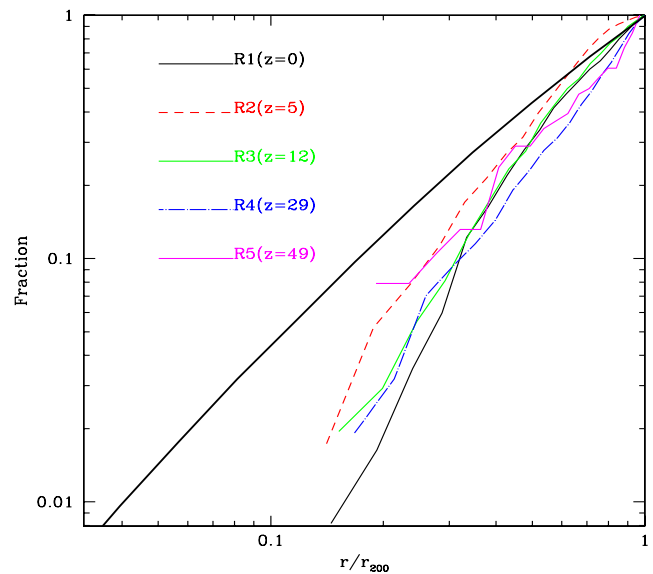,width=250pt,height=250pt}}
\caption{Cumulative radial number density profiles for subhaloes of the most
massive halo at the final time in each of our sequence of resimulations. Only
subhaloes within $r_{200}$ and consisting of at least 30 particles are
considered. The smooth heavy line is the corresponding distribution for
the total mass of the R1 halo.} \label{fig:fig7}
\end{figure}

The radial distribution of subhaloes within the parent halo is less
concentrated than that of the mass as a whole and is quite similar at all
epochs. The cumulative profiles of Fig.~\ref{fig:fig7}) are plotted for
subhaloes with at least 30 particles, the limit to which Gao et
al. (2004a) considered this distribution to be insensitive to resolution
effects; the profiles found here are all similar to those shown in this
earlier paper.

It seems therefore that within $r_{200}$ the internal structure of the most
massive haloes present at high redshift (radial profile, abundance and radial
distribution of subhaloes) differs relatively little from that of present-day
rich cluster haloes, despite the large differences in dimensionless assembly
rate and in the morphology of the structures from which they form.

\section{Large-scale structure at high redshift}
\subsection{Morphology}
The fundamental hypothesis underlying the hierarchical clustering paradigm is
that small objects collapse first and gradually aggregate into larger
systems. This requires that the linear power spectrum of mass density
fluctuations at the beginning of the matter-dominated phase of cosmic
evolution should, in a power-law approximation, be $P(k)\propto k^n$ with
$n>-3$.  At late times when large galaxies, clusters and superclusters are
forming, the effective index of the standard $\Lambda$CDM power spectrum is in
the range $-2<n<0$ and structure growth is indeed hierarchical. At the early
times we are studying here, however, the effective index on the scales
dominating nonlinear growth is very close to $-3$ and an ``orderly''
hierarchical build-up is not guaranteed. This is reflected in the very rapid
early growth of the object we have focussed on, and we may expect the
simultaneous collapse of structure on a very wide range of scales to produce a
different morphology for large-scale structure than is seen at later
times. The image of structure at $z=17$ presented by Yoshida et al. (2003;
their Figure 1) certainly looks quite different from the familiar pictures of
low redshift large-scale structure.

In this subsection we compare large-scale structure at $z=49$ and
at $z=0$. By construction our resimulations are centred on a rare
and massive object. We therefore compare structure in the
high-resolution region of R4 to structure surrounding our massive
cluster in the VLS simulation. It is unclear how the two models
should be scaled to carry out this comparison, since, for any
reasonable definition of the nonlinear mass scale, it is many
orders of magnitude smaller at $z=49$ than at $z=0$. We have
chosen to scale, as in Fig.~\ref{fig:fig3}, by the characteristic
size $r_{200}$ of the central object. Since our $z=49$ object is,
in a dimensionless sense, much rarer than our $z=0$ object (i.e.
farther out on the tail of the mass distribution; see the mass
fraction plot of Figure~\ref{fig:fig1}) one might expect $r_{200}$
for the high-redshift object to be larger than that of an object
which ``correctly'' corresponds to the $z=0$ halo. We would then
underestimate the factor by which early large-scale structure
should be magnified in order to correspond to $z=0$
structure.  In fact, however, the opposite is true: even with the
scaling we have chosen, large-scale structure is {\it much} more
coherent at $z=49$ than at $z=0$.

\begin{figure*}
\hspace{0.13cm}
\resizebox{8cm}{!}{\includegraphics{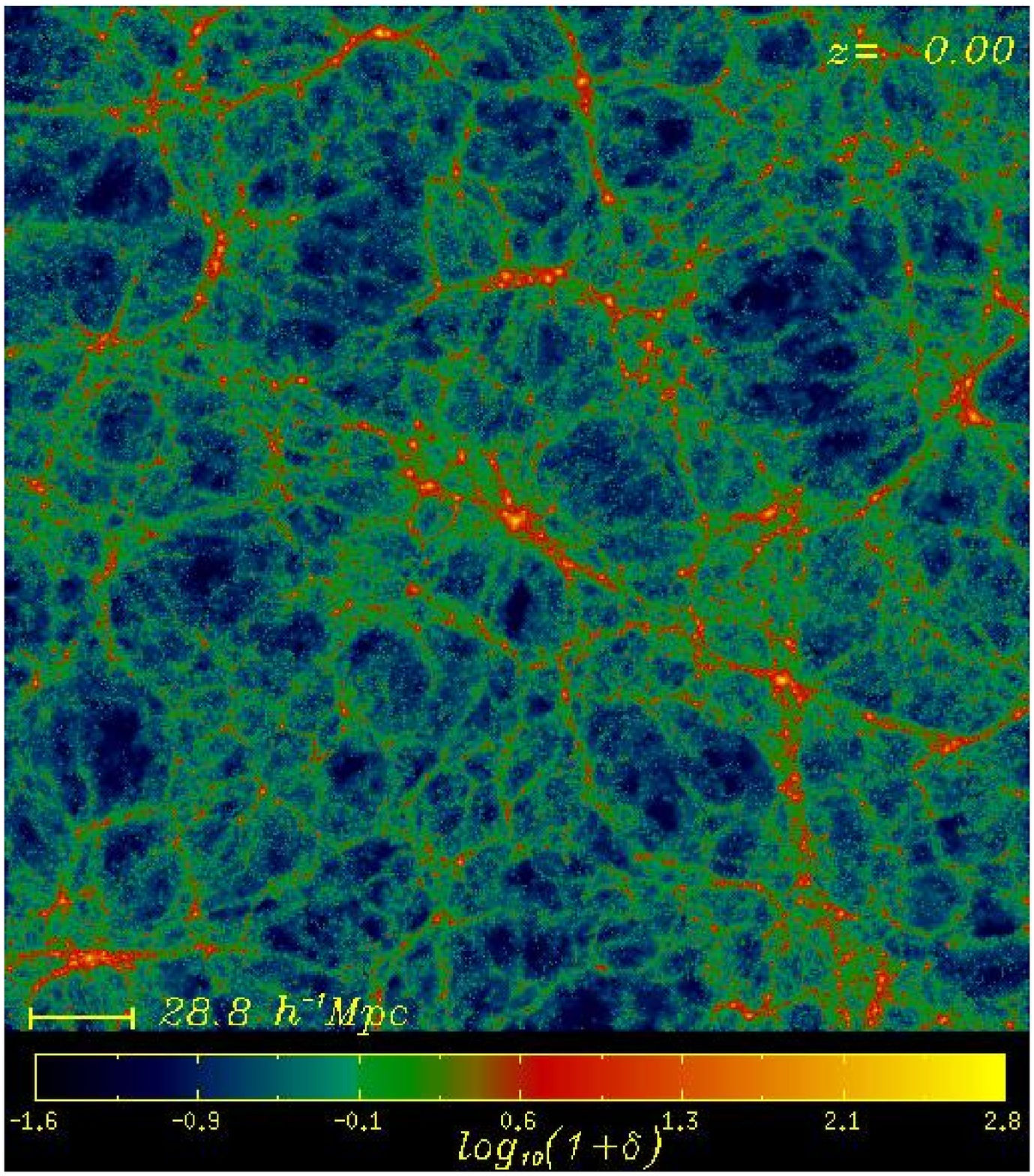}}%
\hspace{0.13cm}\resizebox{8cm}{!}{\includegraphics{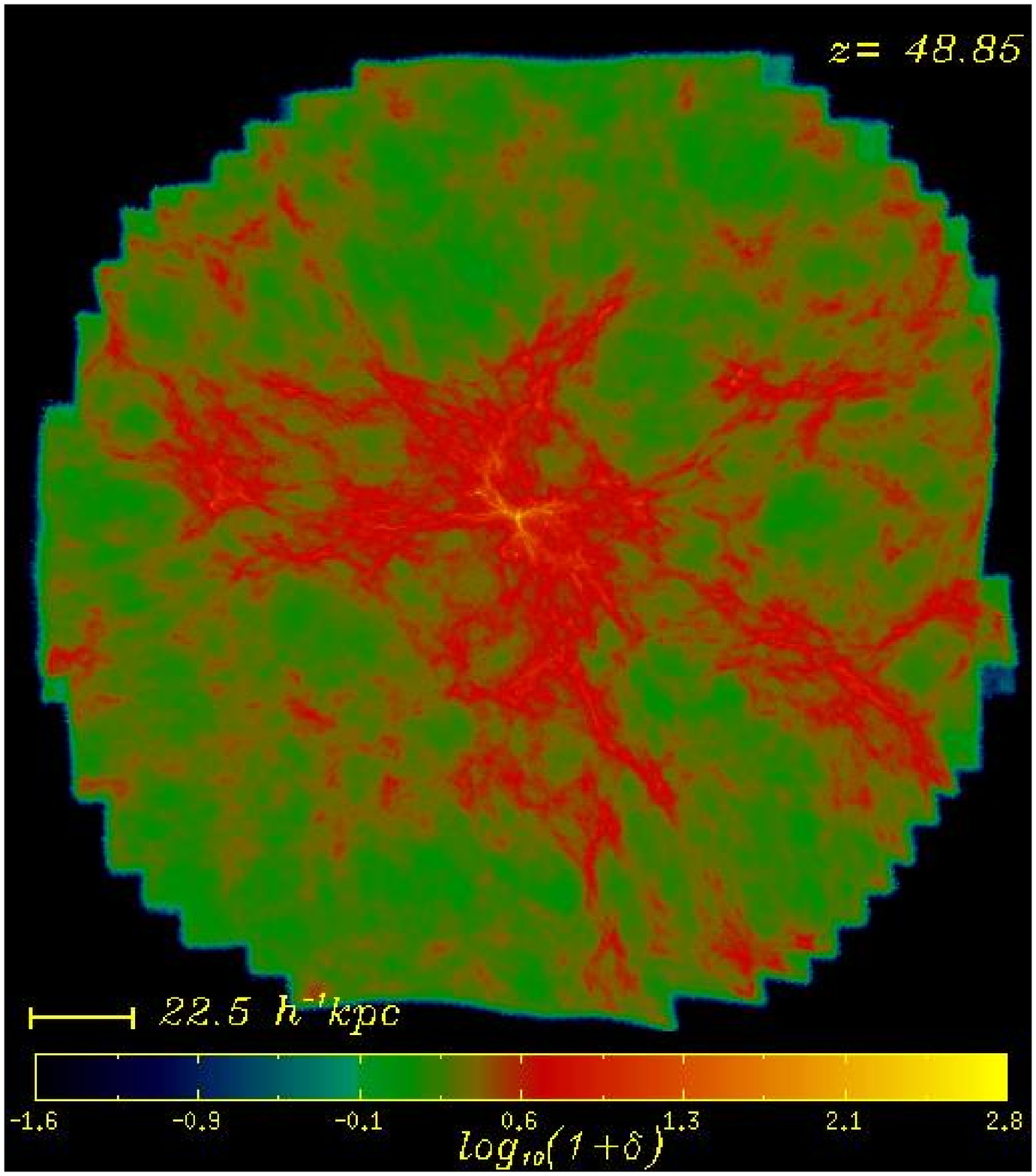}}%
\caption{Projected dark matter density fields at $z=49$ and at
$z=0$ in slices centred on a rare and massive halo. The side of
each plot is $190r_{200}$ and its thickness is $10r_{200}$ where
$r_{200}= 1.2h^{-1}$kpc in comoving units for the central halo at
$z=49$ and $r_{200}=1.5h^{-1}$Mpc for the central halo at $z=0$.
The left-hand plot is taken from the parent VLS simulation and is
centred on the rich cluster halo chosen for our series of
resimulations. The right-hand plot shows the high resolution
region of R4 and is centred on the main halo which was resimulated
in R5. The density fields are normalized in each case by the mean
projected cosmic density at the corresponding epoch and they are
displayed using the same logarithmic colour scale in the two
plots.} \label{fig:fig8}
\end{figure*}

In Fig.~\ref{fig:fig8}, we compare large-scale structure at $z=49$ and
$z=0$ after applying this scaling. The regions shown are $190r_{200}$ on a
side and are $10r_{200}$ in depth. We have normalised by the projected mean
cosmic density at each epoch and used the same logarithmic colour scale to
display the two images.  The qualitative and quantitative differences between
the two plots are dramatic. The mean overdensity across the whole image at
$z=49$ is clearly substantially higher than at $z=0$, and in fact the mean
density of the high resolution region of R4 at the redshift plotted is more
than 2.5 times the global cosmic mean, even though the mass in this region is
five orders of magnitude larger than that of the central halo. In addition,
the large-scale structure at $z=49$ has a single dominant peak and extends
more or less coherently across the whole region, while at $z=0$ there are many
peaks of quite similar mass and morphology to the central one and typical
filaments and voids are smaller than the region plotted.  Clearly it is
dangerous to assume that large-scale structure in the early universe is just a
scaled version of that familiar from simulations of low redshift structure
formation.

\subsection{Biases in the abundance of haloes}

\begin{figure*}
\resizebox{8cm}{!}{\includegraphics{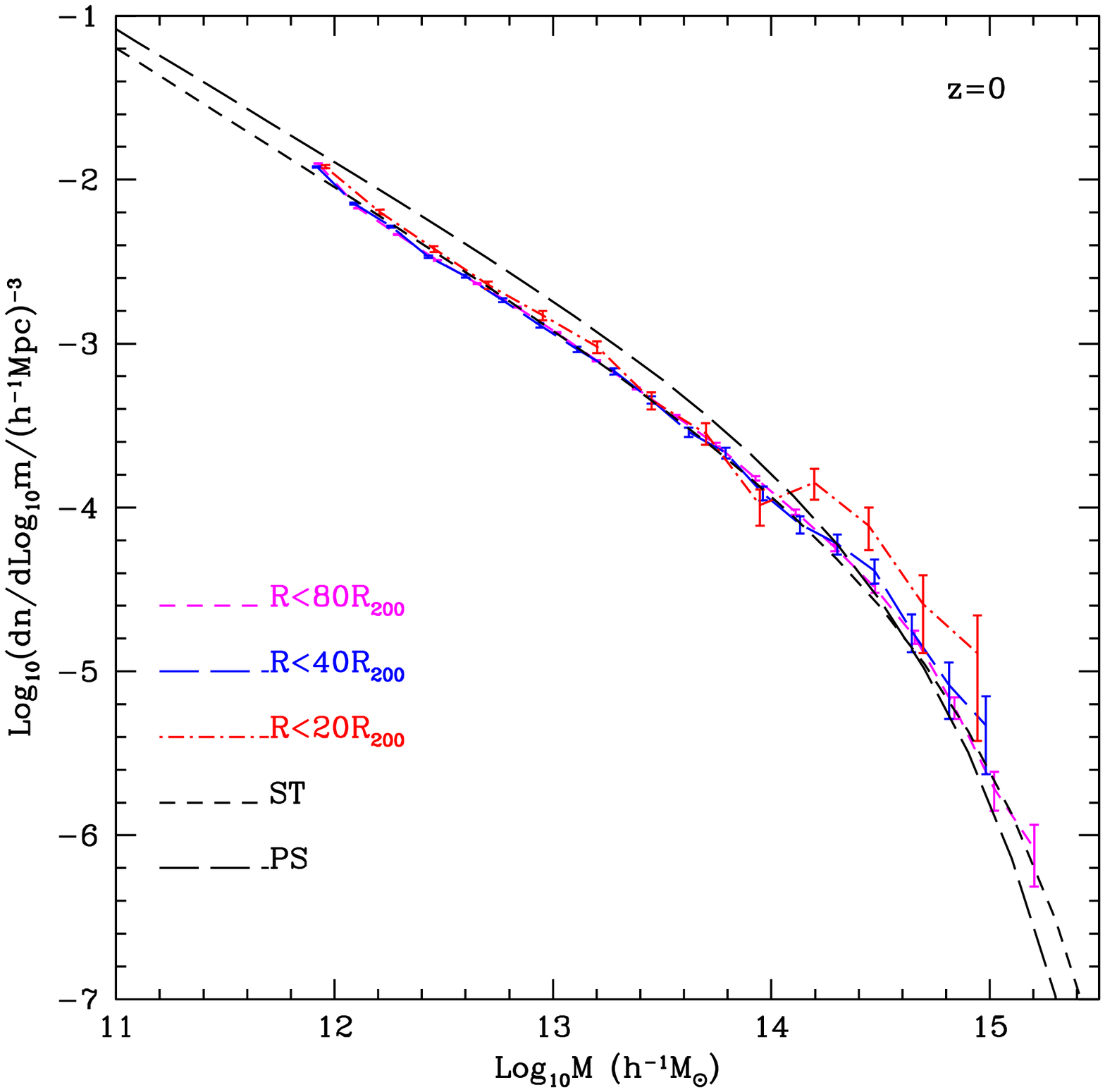}}
\hspace{0.0cm}\resizebox{8cm}{!}{\includegraphics{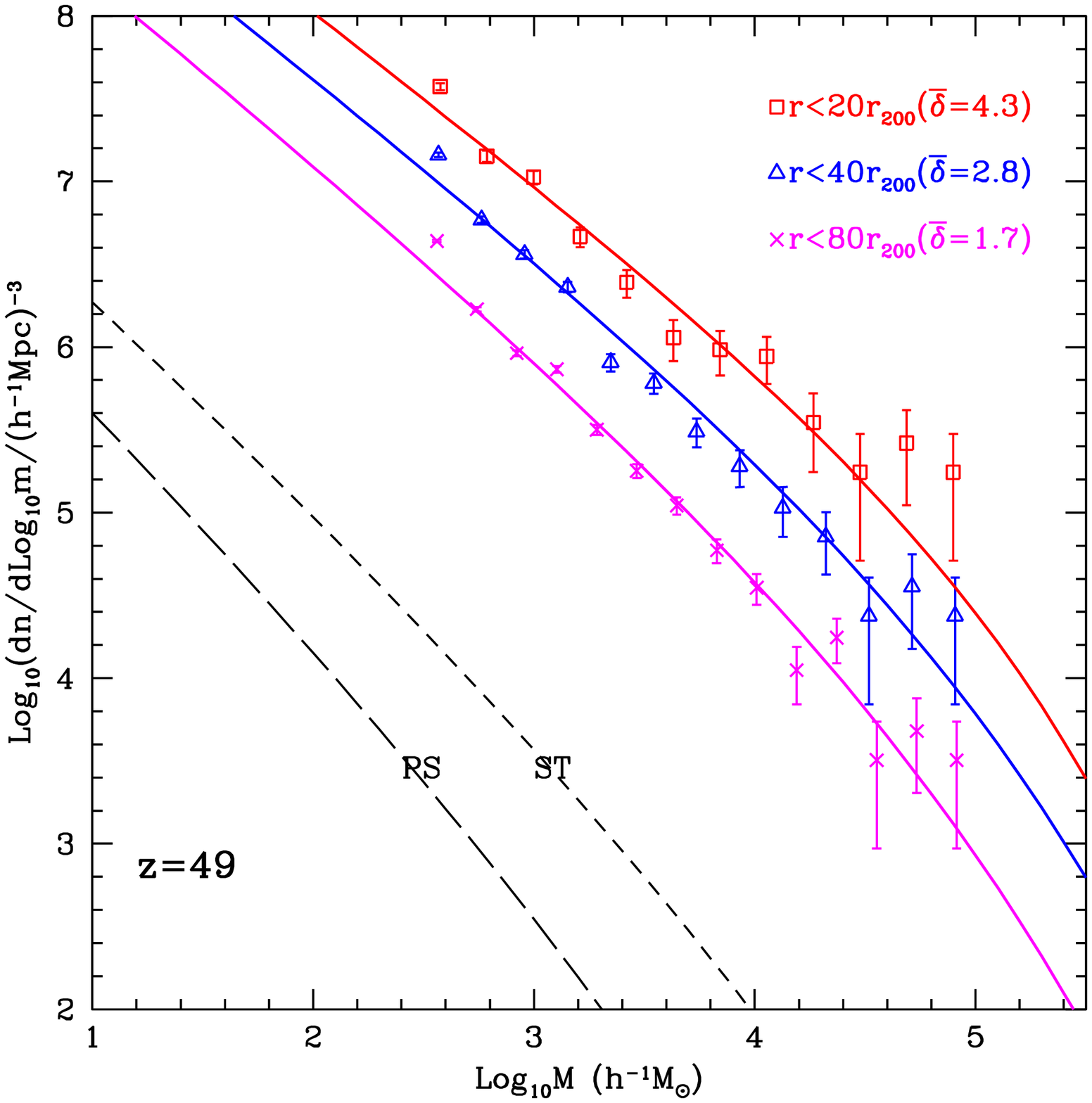}}
\hspace*{1.36cm}\ \\
\caption{Halo mass functions in the VLS (left) and R4 (right)
simulations for spherical regions of radii $80r_{200}$,
$40r_{200}$ and $20r_{200}$ centred on the R1 halo at $z=0$ and on
the R5 halo at $z=49$. Objects were identified using the SO(180)
algorithm. The error bars represent Poisson uncertainties in the
counts. In both panels we give predictions based on the PS and ST
formulae for the halo abundance in typical regions of the
Universe. In the right-hand panel, solid lines show predictions of
the extended Press--Schechter model for regions with the measured
nonlinear overdensities.} \label{fig:fig9}
\end{figure*}

The regions followed by our resimulations are not typical regions
of the universe, since, by construction, they are centred on a
rare and massive object. Indeed, all the material followed at high
resolution in R$n$ falls onto this object during the evolution of
R$(n-1)$ for $n=2,3,4$ and $5$.  It is this bias which is
responsible, of course, for the substantial overdensity of the
whole $z=49$ region plotted in Fig.~\ref{fig:fig8}, and it results
in accelerated growth of objects throughout the overdense region
in comparison with ``typical'' regions of the universe.  This may
be quantified by comparing the mass function of haloes in the high
resolution regions of our resimulations with that expected at the
same redshift for the universe as a whole. Fig.~\ref{fig:fig9}
shows, as a function of mass, the abundance of objects identified
by the SO(180) algorithm in spherical regions centred on the most
massive halo at two different redshifts. The left-hand plot is
based on the VLS simulation at $z=0$ and shows results for spheres
of radius $80$, $40$ and $20$ times $r_{200}= 1.5h^{-1}$Mpc.  The
right-hand plot is based on R4 at $z=49$ and shows results for
spheres with the ``same'' radii, except that now
$r_{200}=1.2h^{-1}$kpc in comoving units. These plots thus
correspond to the simulations and redshifts illustrated in
Fig.~\ref{fig:fig8}.

Both panels of Fig.~\ref{fig:fig9} also show predictions for the halo mass
function in ``typical'' regions of the universe based on the standard
Press-Schechter (PS; Press \&~Schechter 1974) and Sheth \& Tormen (ST; Sheth
\&~ Tormen 1999) formulae. At $z=0$ these formulae agree moderately well and
the ST prediction, in particular, is very close to the fitting function
proposed by Jenkins et al. (2001) as a good representation of the best
available simulation data. At $z=49$, however, the two predictions disagree
badly -- by a factor of about 10 at $1000h^{-1}{\rm M}_\odot$ and by a factor
of about 100 at $10^5h^{-1}{\rm M}_\odot$. In this regime there are no
reliable simulations with which they can be compared, so it is unclear which
(if either) is most correct.  For the largest high-resolution simulation of a
``typical'' region carried out so far, the so-called Millennium Run, the ST
formulae continue to fit the high mass tail of the concordance $\Lambda$CDM
mass function out to $z=10$ where the PS formulae already underpredict by
about an order of magnitude (Springel et al. 2005). Nevertheless, these
results are still a factor of 5 in redshift and $10^6$ in mass from the regime
of concern here.

Both the PS and the ST formula fit the $z=0$ data well for spheres of radius
$80$ and $40r_{200}$ and the simulation results in the two cases are
indistinguishable.  For the sphere of radius $20r_{200}$ , the simulation
results are similar at small mass but appear high at the large mass end.
Within the two larger spheres the mean overdensity is close to zero, so no
substantial bias is expected theoretically.  The situation is dramatically
different at redshift $z=49$. Even though the ST prediction is much larger
than that from PS theory, it still lies far below any of the mass functions
measured from the simulation.  In addition, the three spheres produce mass
functions which differ considerably, with the smallest sphere having the
largest abundances.  As noted above, the overdensities within these spheres
are substantial, and as we now show, their mass functions can be predicted
surprisingly well by inserting these overdensities into a suitable version of
the conditional or extended Press-Schechter model (EPS; Bond et al. 1991;
Bower 1991; Lacey \& Cole 1993, 1994; Mo \& White 1996).

According to the EPS model as worked out by Mo \& White (1996), the fraction
of the material in a large spherical region of total mass
$M_0$ and linear overdensity extrapolated to the present day $\delta_0$
which, at redshift $z$, is contained in dark haloes with mass in $(M_1,
M_1+dM_1)$, is given by
\begin{eqnarray}\label{EPS}
\lefteqn{ f(M_1,z|M_0,\delta_0)dM_1=
                          (\frac{1}{2\pi})^{1/2}\frac{\delta_1-\delta_0}
                          {(\Delta_1^2-\Delta_0^2)^{3/2}}} \nonumber\\
                          & & \exp[-\frac{(\delta_1-\delta_0)^2}
                          {2(\Delta_1^2-\Delta_0^2)}]\frac{d\Delta_1^2}{dM_1}dM_1,
\end{eqnarray}
where $\Delta_0$ and $\Delta_1$ are the {\it rms} linear fluctuations
extrapolated to the present day in spheres which on average contain masses
$M_0$ and $M_1$ respectively, and $\delta_1$ is the critical linear overdensity
for collapse at $z$ also extrapolated to the present day.  To evaluate
$\delta_0$, we must relate the nonlinear overdensity $\delta$ measured at
redshift $z$ in Eulerian space to the original linear overdensity in
Lagrangian space. Based upon the spherical collapse model, Mo \&
White(1996) provided an analytical fitting formula which links these two
overdensities in an Einstein-de Sitter universe. Sheth \& Tormen (2002)
later showed that this formula remains reasonably accurate for all cosmologies
of interest.
\begin{eqnarray}{\label{nonden}}
\delta_0(\delta,z)\!\!\! &=&\!\!\! {\delta_1\over
1.68647}
 \times \Biggl[1.68647 - {1.35\over (1+\delta)^{2/3}} \nonumber\\
&&\qquad\qquad\quad - {1.12431\over (1+\delta)^{1/2}}
                  + {0.78785\over (1+\delta)^{0.58661}}\Biggr]
\label{d0mow}
\end{eqnarray}

\begin{figure}
\resizebox{8cm}{!}{\includegraphics{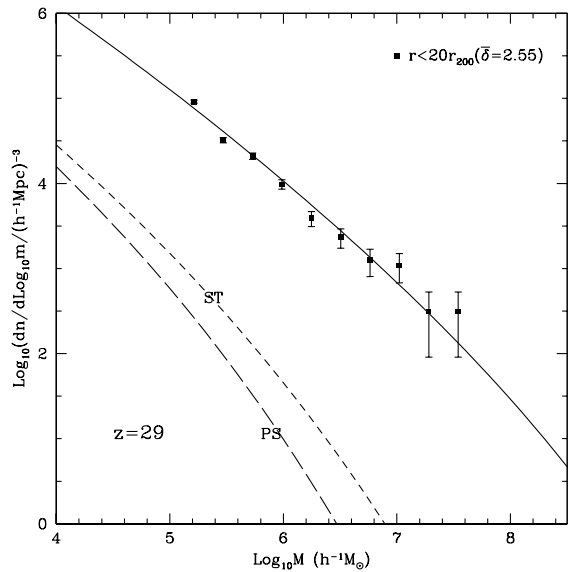}}
\caption{Differential mass function in a spherical region with radius
$20r_{200}$ centred on the most massive halo in our R3 simulation at
$z=30$. The solid line shows the prediction from EPS theory. The error bars
assume Poisson uncertainties in the counts.} \label{fig:fig10}
\end{figure}

The nonlinear overdensity in the spherical regions used to make Figure
\ref{fig:fig9} can be estimated straightforwardly from the
simulations. For R4 at $z=49$ we find $\delta_{nl}=1.7, 2.8$ and $4.3$
for spheres of radius $80r_{200}$, $40r_{200}$ and $20r_{200}$
respectively. For the VLS simulation at $z=0$ the corresponding
numbers are -0.02, 0.17 and 0.49. These numbers can be converted to
linear overdensities using equations (\ref{nonden}), while the radii of the
spheres can be converted to Lagrangian radii simply by multiplying by
$(1+\delta_{nl})^{1/3}$. Equations (\ref{EPS}) then predict the halo
abundance in these regions. We plot the results as thick solid lines
in the left panel of Fig.~\ref{fig:fig9}. (We do not show these
curves at $z=0$ since they differ little from the standard PS prediction.)
Clearly, the agreement with the measured abundances is very
good. Fig.~\ref{fig:fig10} shows that the EPS model works equally well
at $z=30$. For this plot we consider the halo abundance in a spherical
region of radius $20r_{200}$, the largest sphere that fits completely
within the high resolution region of R3 at this redshift.  As noted
above, tests of the standard PS formula show increasingly poor fits to
the high mass tail of the global mass function at high redshift (and, in
particular, much worse fits than the ST formula), so it is remarkable
that the EPS mass function gives a such an accurate estimate of halo
abundance in high density regions at these early times. The redshift
and mass ranges tested here, $z=30 - 50$ and $M\sim 10^5 - 10^6{\rm
M_\odot}$, are those most relevant for the collapse of the haloes which
host the first stars.

The EPS formalism may also be used to calculate the growth history
of a given halo. At given (high) redshift, we estimate the
expected mass of the most massive progenitor of the final halo by
integrating the EPS progenitor mass function down from infinity
until the expected total number of progenitors is equal to $1$. We
then adopt the mean progenitor mass over this range as our
estimate for the largest progenitor mass. The thick line in
the upper panel of Fig.~\ref{fig:fig1} shows the resulting prediction
for the growth in mass of the dominant halo in our resimulation
series. (Recall that the final object is not the most massive halo at
$z=0$ in R1 but rather a nearby halo of mass $1.07 \times
10^{14}h^{-1}{\rm M_\odot}$.) Remarkably, the theoretical
prediction follows our simulation results very closely, with at
most a $10$ per cent shift in redshift at early times. We believe
this close agreement to be at least partly fortuitous, since we
have also resimulated the most massive $z=5$ progenitor of the
rich cluster halo we originally picked from the VLS simulation.
Although at $z=5$ it is only 15\% less massive than the object we
have concentrated on in this paper (and at $z=0$ is a factor $8$
{\it more} massive), by $z=29$, the earliest time for which we
currently have a reliable mass estimate, it is less massive than
our principal object by a factor of $10$. Clearly, there is
substantial scatter in growth histories of the kind we are
studying.

\section{Implications for the formation of the first stars}
In the context of the $\Lambda$CDM cosmogony, understanding the
formation and evolution of early dark haloes is a prerequisite for
understanding the formation of the first stars. Although first
star formation depends strongly on additional hydrodynamic and
radiative processes, dark halo collapse provides the dynamical
context for these processes, and so can serve as a useful guide to
when and where the first stars may form.  We already saw in
Fig.~\ref{fig:fig1} and Fig.~\ref{fig:fig4} that a halo of virial
temperature $\sim 2000$K is in place in our simulation series by
$z=49$. These are the properties usually suggested as appropriate
for the condensation of the first stellar generation through
molecular hydrogen cooling. By $z=39$ the biggest halo has high
enough temperature for efficient cooling by atomic processes. We
now briefly consider the implied abundance and clustering of such
objects. A more detailed discussion is given in our companion
paper (Reed et al. 2005c).

The lower right-hand panel of Figure~\ref{fig:fig1} shows the ST
formulae to predict that at $z=49$ the mean abundance of objects
at least as massive as our dominant halo is $\sim
10^{-1}h^3$Mpc$^{-3}$, while at $z=39$ it is $\sim
10^{-2}h^3$Mpc$^{-3}$. These numbers correspond roughly to the
present-day abundance of $10^{11}$ and $10^{12}h^{-1}{\rm
M_\odot}$ haloes respectively. The lower left-hand panel of
Figure~\ref{fig:fig1} demonstrates that only a tiny fraction of
all mass is in such haloes, about $2\times 10^{-7}$ at $z=49$ and
about $3\times 10^{-6}$ at $z=39$. Since the masses of the first
stars are expected to be a few hundred M$_\odot$, the fraction of
all baryons in stars at $z=49$ is predicted to be around
$10^{-10}$. This is far too low for their UV flux to reionise a
significant fraction of the IGM. These early objects would thus
create isolated HII regions with a very small filling factor.
Atomic cooling might allow more efficient condensation of the
baryons into stars in the $z=39$ objects, but even then the
abundance is too low for the HII regions to percolate.  In
addition, activity related to a central star formed at $z\sim 49$
might eject the baryons from these objects before their virial
temperature reaches the critical value for atomic cooling.

Closely related to the fact that the main progenitors of high-mass
objects accrete matter very rapidly at high redshift is the fact
that the abundance of objects above a given (high) mass or
temperature threshold increases very rapidly with time. Thus while
the ST prediction for the mass fraction in objects with virial
temperature above 2000K is $2\times 10^{-7}$ at $z=49$, it has
already increased to $3 \times 10^{-5}$ by $z=39$ with a corresponding
abundance of $\sim 10h^3$Mpc$^{-3}$, well above that of dwarf galaxies
today. Note that these strong dependences are also reflected in
considerable sensitivity of these ST predictions to the particular
form used for the $\Lambda$CDM power spectrum. Here, as in our
simulations, we have used the spectrum given by {\small CMBFAST}
(Seljak \& Zaldarriaga 1996). In the regime we are considering,
switching to the analytical fits quoted by Bond \& Efstathiou
(1987) or Bardeen et al. (1986) alters the abundance at fixed halo
mass and redshift by up to an order of magnitude, and the redshift
at fixed halo mass and abundance by up to $\Delta z\sim 5$.

\begin{figure}
\resizebox{8.0cm}{!}{\includegraphics{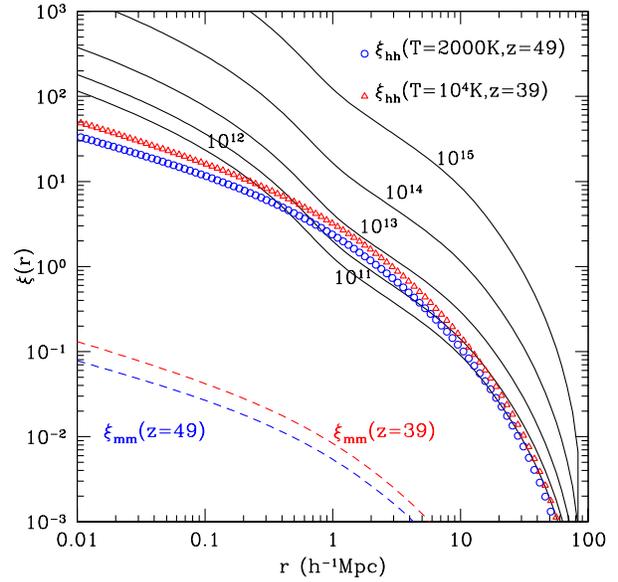}}\\
\caption{Halo auto-correlation functions. Open circles show $\xi_{hh}$ for
haloes of virial temperature 2000K at $z=49$, while open triangles refer to
10000K haloes at $z=39$. Solid lines are predictions for $z=0$ haloes with
the masses noted.  All these functions were calculated using the procedures of
Mo \& White (1996). For comparison, the dashed lines show the correlation
functions predicted for the mass distribution at $z=49$ and $z=39$.}
\label{fig:fig11}
\end{figure}

A final point of interest for the first stars is the spatial clustering of
their potential host haloes. This has a direct impact on the likelihood that a
star or star cluster can affect star formation in neighbouring haloes. It also
has consequences for the morphology of the IGM during its reionisation and so
for the form of the predicted signal in future redshifted 21cm surveys
(e.g. Ciardi \& Madau 2003). The halo autocorrelation function is easily
estimated using the formalism proposed by Mo \& White (1996) which is based
upon the EPS and top-hat collapse models and was tested by them against direct
N-body simulations. The halo auto-correlation function, $\xi_{hh}(r)$, is
given by

\begin{equation}\label{xibias}
\xi_{hh}(r,M,z)=b^2(M,z)\xi_{mm}(r,z)~.
\end{equation}
where
\begin{equation}
b(M,z)=1+\frac{\nu^2(M,z)-1}{\delta_c(z)},~~~\nu = \delta_c(z)/\Delta(M),
\end{equation}
is the bias factor for haloes of mass $M$ at redshift $z$ (Cole \&~Kaiser
1989; Mo \&~White 1996, 2002; Jing 1998), and $\xi_{mm}$ is the {\it
non}linear auto-correlation function of the underlying mass, estimated to
sufficient accuracy from the linear power spectrum by the procedure of Peacock
\& Dodds (1996). Equation \ref{xibias} is only valid on scales significantly
larger than the size of the haloes, and, like the PS and ST formulae, it has
not been tested numerically in the regime where we use it here. In
Fig.~\ref{fig:fig11}, we compare the correlation functions for 2000K haloes at
$z=49$ and for 10000K haloes at $z=39$ with those for haloes with a variety of
masses today. (Note that lengths in this plot are in comoving units.) The halo
correlation amplitudes at these early times are several hundred times that of
the underlying mass. On scales of a few Mpc they are comparable to that of
$M\sim 10^{12}h^{-1}{\rm M_\odot}$ dark haloes today. The correlation length
(defined by $\xi(r_0)=1$) is $r_0=2.5h^{-1}$Mpc for our $z=49$ haloes and is
$r_0\sim 3h^{-1}$Mpc for the $z=39$ haloes. Thus on Mpc scales, the
comoving clustering pattern of the first stars and galaxies may be quite
similar to that of dwarf galaxies today.

\section{Conclusions}

We have invented a novel strategy which allows us to follow the
build-up of a massive object within the concordance $\Lambda$CDM
cosmogony over an unprecedentedly large range in redshift and
mass. In the particular resimulation series presented in this
paper we track the growth of one particular object from a mass of
about $10h^{-1}{\rm M_\odot}$ at $z=80$ to about
$10^{14}h^{-1}{\rm M_\odot}$ at the present day.  Throughout the
series the full cosmological environment of the object is
correctly represented so we can study evolution both of its
internal structure and of the ``large-scale'' structure in which
it is embedded. As a by-product we are able to estimate
approximately when the baryons associated with this particular
object would first have been able to collapse to form a star. Our
main results may be summarized as follows:

\begin{description}

\item[(1)] Massive objects at early times accrete mass extremely rapidly and
form in regions where the overdensity is high on much larger scales than those
of the objects themselves.

\item[(2)] At early times ($z\ge 12$) the filamentary structure surrounding
the massive objects is much stronger than at lower redshifts ($z\le 5$).

\item[(3)] At very high redshift ($z\sim 50$) large-scale structure is
qualitatively different from that in the low redshift universe. In
particular the characteristic size of coherent structures is much larger
in relation to the typical size of collapsed objects.

\item[(4)] Despite this, the internal structure of massive early dark
haloes is quite similar to that to their present-day counterparts, both in
terms of density profiles and in terms of substructure. Although
early objects are less concentrated and have less substructure, the
differences are relatively small.

\item[(5)] The number density of haloes in overdense regions at
high redshift is surprisingly well described by the extended
Press-Schechter model, at least in the redshift interval $z=50 -
30$ that we study here.

\item[(6)] By $z=49$ our main object has reached virial temperature $2000$K
and should soon form sufficient molecular hydrogen for gas to condense into a
central massive star.  Pockets of star formation may thus have appeared much
earlier than previously thought.  At $z=49$ such 2000K haloes have a comoving
number density comparable to that of $\sim 10^{11}h^{-1}{\rm M_\odot}$ haloes
today, but by $z=40$ this has already increased by almost two orders of
magnitude.  By $z=39$ our main halo has a virial temperature of 10000K so its
gas can start cooling by atomic processes and turning, perhaps, much more
efficiently into stars.

\item[(7)] Over the redshift range $30<z<50$ haloes like the one
we have simulated have comoving correlation lengths up to
$3h^{-1}$Mpc, similar to those of present-day dwarf galaxies. The
ionisation and dissociation structures their associated stars
produce is studied in our companion paper (Reed et al. 2005c).
\end{description}

By clarifying how the dark matter distribution evolves in and around extreme
objects, our simulations set the scene for the more complex processes of gas
cooling, collapse and fragmentation that result in the first stars. We will
study these processes in forthcoming papers using direct simulation of the
baryonic component.

\section*{Acknowledgements}
The simulations used in this paper were carried out on the IBM Regatta
supercomputer at the Computing Center of the Max-Planck-Society in
Garching, Germany and on the Sun Microsystems Cosmology Machine at the
Institute for Computational Cosmology of the University of Durham,
England. GL thanks Cedric Lacey and Yipeng Jing for useful
discussions. We also thank the referee for a careful reading of our
paper and a helpful report.

\label{lastpage}

\end{document}